\documentclass[aps,prd,reprint,superscriptaddress, floatfix]{revtex4-1}

\usepackage{color}
\usepackage{amssymb}
\usepackage[normalem]{ulem}
\usepackage{graphicx,color}
\usepackage{amsmath}
\usepackage{subfigure}
\usepackage[normalem]{ulem}
\usepackage{booktabs}
\usepackage{xcolor}
\usepackage{epsfig,graphics,color,graphicx,amsmath}

\colorlet{darkgreen}{green!50!black}
\colorlet{brightyellow}{yellow!75!red}
\colorlet{orange}{red!50!yellow}
\colorlet{darkblue}{blue!60!black}
\colorlet{darkred}{red!80!black}

\usepackage{soul}

\usepackage{mathtools}
\usepackage{mathrsfs}
\usepackage{bm}
\usepackage{relsize}
\usepackage{indentfirst}

\usepackage{xcolor}

\usepackage{amssymb,amsmath,multirow,epsfig,graphicx,color}

\usepackage{epsfig}
\usepackage{graphicx,amsmath}
\def\be{\begin{eqnarray} &&}
\def\nonu{\nonumber \\ &&}

\def\ee{\end{eqnarray}}
\def\psla{\slash \! \! \!}

\def\psla{  \slash \!  \!\!}

\begin{document}

\title{ Observing the Minkowskian dynamics of the pion on the null-plane}

\author{W. de Paula}
\affiliation{Instituto Tecnol\'ogico de Aeron\'autica,  DCTA, 
12228-900 S\~ao Jos\'e dos Campos,~Brazil}
\author{E.~Ydrefors}
\affiliation{Instituto Tecnol\'ogico de Aeron\'autica,  DCTA, 
12228-900 S\~ao Jos\'e dos Campos,~Brazil}
\author{J.H.~Alvarenga Nogueira}
\affiliation{Instituto Tecnol\'ogico de Aeron\'autica,  DCTA, 
12228-900 S\~ao Jos\'e dos Campos,~Brazil}
\affiliation{Dipartimento di Fisica, Universit\`a di Roma La Sapienza"
 P.le A. Moro 2, 00185 Rome, Italy}
 \affiliation{
INFN, Sezione di Roma, P.le A. Moro 2, 00185 Rome, Italy}
\author{T.~Frederico}
\affiliation{Instituto Tecnol\'ogico de Aeron\'autica,  DCTA, 
12228-900 S\~ao Jos\'e dos Campos,~Brazil}
\author{G. Salm\`e}
\affiliation{
INFN, Sezione di Roma, P.le A. Moro 2, 00185 Rome, Italy}

\date{\today}

\begin{abstract}
A dynamical model  is applied to the study of the pion valence light-front wave 
 function, obtained from the  actual solution of the Bethe-Salpeter equation in Minkowski space, resorting to the Nakanishi integral representation. The kernel is simplified to a ladder approximation containing constituent quarks,
an effective massive gluon exchange, and the scale  of the extended quark-gluon interaction vertex.
  These three input parameters  carry the infrared 
  scale $\Lambda_{QCD}$ 
   and are   fine-tuned to reproduce the pion weak decay constant, within a range
   suggested by lattice calculations. Besides $f_\pi$, we present and discuss other interesting 
   quantities on the null-plane, like:  (i) 
    the valence probability,
       (ii) the dynamical functions depending upon the  longitudinal or the  transverse
   components of the light-front (LF)  momentum,  represented by 
    LF-momentum distributions and  distribution amplitudes, and 
    (iii) the 
    probability densities both in the LF-momentum space and 
    the 3D space given by the Cartesian product of the 
     covariant Ioffe-time and
    transverse coordinates, in order to perform  an analysis of the dynamical features in a
    complementary way. The proposed analysis of the Minkowskian dynamics inside the pion, though 
    carried out at the initial stage,
     qualifies the Nakanishi integral
    representation as an appealing effective tool, with still unexplored potentialities
     to be exploited for
    addressing correlations between dynamics and  observable properties.  

   \end{abstract}
\maketitle

\section{ Introduction.}

The pion 
{plays a pivotal role  within
 Quantum Chromodynamics (QCD), since its Goldstone boson nature is }
associated with the dynamical generation of the mass of the hadrons and nuclei constituting the visible universe. 
The pion has a rich  structure that stems from
the spin degrees of freedom of its constituents, which are necessarily associated with the covariant Minkowski space 
formulation of QCD. Its dynamics entangles in a conspicuous way the space 
 and spin  distributions of the fundamental degrees of freedom within a hadron.  
 The pion is still puzzling by its Goldstone boson nature and its composition in terms of quarks and gluons~
  (see, e.g., Ref.~\cite{Aguilar:2019teb}), 
 so that  its momentum distributions,  the most typical
 dynamical quantities,  have been a target of intense investigation in recent years~\cite{Barry:2018ort,Shi:2018zqd,Bednar:2018mtf,Oehm19,Lan:2019vui,Lan:2019rba,DingPRD2020,DingCPC2020,Sufian:2020vzb,Chang:2020kjj},  as well as of planned  experimental research at  the 
 future Electron Ion Collider.

 Hadron imaging is driving experimental \cite{Dudek2012,adolph2013hadron} and theoretical \cite{Accardi:2017pmi} research efforts  
 towards the exploration of 
 the Fock-space   structure of the  light-front (LF) wave functions, even beyond the valence component. 
By properly selecting the imaging space, the  Fock-space structure of the hadron  can be revealed by looking at single-parton distributions 
 with or without spin polarization, double-parton distributions 
(see, e.g., Refs.~\cite{Rinaldi:2016jvu,Kasemets:2017vyh}), 
 triple-parton distributions and in general $n$-parton distributions. 
 It is clear that such a program, when developed, would provide the ideal framework to study in great detail  the
 Fock-space components of the hadron wave function. Therefore, research efforts  to explore
 the  Fock-space content of the hadron  LF wave function are necessary, 
 either by using Euclidean 
 Lattice discretization, i.e. lattice QCD (LQCD) (see, e.g., Ref. \cite{Beane2011}) or 
 continuous QCD techniques (see, e.g., Ref.~\cite{CloPPNP14}).
 
 The challenge on the theory side is to extract from Euclidean calculations the relevant 
 observables defined in the Minkowski space. Vigorous research is pursued to 
 extend LQCD calculations, carried out in Euclidean space,   
  and eventually attain  the   parton distribution functions (PDFs). With such an
 aim, several strategies have been proposed, like the one based on (i) the quasi-parton distribution functions 
  (QPDFs)  \cite{Ji:1996ek} (see, e.g., 
Ref.~\cite{Alexandrou:2017huk} for early results), (ii) the pseudo-parton
distribution functions (PPDFs) \cite{Radyushkin:2017cyf} (see, e.g., Ref.
\cite{Joo:2019bzr} for the  pion case)
and (iii)  the so-called {\em lattice cross-section} method, as applied 
 in  Refs. \cite{Sufian:2019bol,Sufian:2020vzb}.  
 Another approach is the analytical continuation  of the  solution of the Euclidean Bethe-Salpeter (BS) equation 
  to the Minkowski space. This method uses the Nakanishi integral representation (NIR) \cite{Nakanishi:1971} 
 of the Euclidean BS amplitude, 
 allowing to perform the analytical extension to the Minkowski space.
 Despite that the extraction of  the Nakanishi weight-function
from the Euclidean calculations constitutes  an ill-posed numerical problem 
and   is highly challenging, the NIR method   has been applied for obtaining the pion valence PDF from
 the Euclidean amplitude \cite{LeiPRL13} and the idea was further explored in Refs.
  \cite{CloPRL13,ChanPRL13}.
Although delicate issues on non perturbative renormalization were pointed out 
in Ref.~\cite{Rossi:2017muf},  there is a possibility of using the NIR to compute QPDFs, allowing to 
bridge the continuum Minkowski-space QCD to LQCD calculations.  

 As is well-known \cite{Brodsky:1997de}, the {\em phenomenological} description of hadron states on the null-plane
 can take 
advantage of a meaningful Fock expansion, once  a tiny  mass is assigned 
to the exchanged bosons. Hence the LF approach  can usefully exploit the powerful physical intuition
based on the Fock space, without the difficulties present in the covariant
phenomenological models. In order to formally obtain  the LF valence wave function from 
 the BS amplitude, one has to  perform its 
 projection onto the null-plane.
 As a matter of fact, by applying the LF projection
 to the 
 correlator, built with the minimum number of field operators and providing 
a non-zero matrix element between the vacuum and the hadron state
 \cite{Sales:1999ec,Sales:2001gk}, one eventually gets the valence 
 wave function (see also, e.g.,  Ref. \cite{FSV1} and Appendix \ref{app_val}). Indeed,
 one could generalize the procedure, since 
 to a given hadron state one can  associate an infinite number of BS amplitudes, 
 with any number of legs,  i.e. 
 quarks and gluons, compatible 
with the hadron quantum numbers.  In turn,  each BS amplitude,
when projected onto 
the null-plane,  gives the  corresponding amplitude of the Fock
state with the number of constituents equal to the fields present in the BS amplitude
itself. In principle,   images of the probability densities obtained from 
those LF amplitudes shed light on the dynamics inside the hadron with an
unprecedented level of detail on the Fock content of the hadron state.  However,
even the BS amplitude with the minimal number of legs, has information on the full LF 
Fock-space composition of the hadron \cite{Sales:1999ec,Sales:2001gk}.
It should be recalled that gauge links are always required between the quark field operators in an observable 
\cite{Collins:2011zzd}, like e.g., 
a photon absorption amplitude, to keep color gauge invariance, while the BS amplitude by itself is not 
gauge invariant. 
 
On the theory side, the NIR of the BS amplitude 
can be a useful tool to solve  BS equations in Minkowski 
space~\cite{Kusaka:1995za,Kusaka:1997xd,Karmanov:2005nv,FSV2,FSV3,dePaula:2016oct,dePaula:2017ikc,AlvarengaNogueira:2019zcs}, 
and provide the parton 
distribution amplitudes as well as the elementary fragmentation functions. 
We have no proof yet that QCD, which embodies confinement, allows such integral 
representation in the non perturbative domain, although 
bound state solutions of the BS equation (without confinement) can be  
actually achieved by using NIR   (more precisely, by formally converting the
BSE into  a  generalized
eigenvalue problem). 
 In order to use the integral 
representation to solve the BS equation in its full glory, it is necessary to write 
the quantities entering the kernel,  e.g. 
 the quark-gluon vertex and the propagators,  in terms of the NIR. As we mentioned, the nice feature of the
 NIR is that one can analytically extend 
it from Minkowski to  Euclidean space by performing the Wick-rotation, 
 so that a direct comparison 
with LQCD results can be feasible. 
It is worthwhile to point out that  within the  NIR approach one can prove that the form of 
the valence wave function
for asymptotically large transverse-momentum presents the factorization  of the dependences    upon 
$\xi$ (the fraction of the longitudinal momentum) and  $k_\perp$, naturally recovering the
 power-law fall-off in the UV region 
 \cite{Gutierrez:2016ixt,dePaula:2016oct,gigante2017bound}.
 Such a property can be
extended to higher Fock components of the wave function. 
Furthermore,  at the initial scale, even the simplified calculation of the PDF, 
  based on the Mandelstam
  formula involving the BS amplitude, will include partons
   from Fock-state components beyond the valence one. This is an immediate consequence of the solution of the BS equation in
 Minkowski space.

In the perspective of exploring dynamical models,  incorporating as much as possible 
  non perturbative features of QCD
 in Minkowski space,  
 and to take advantage of
 the results for    building useful hadron  imaging, 
 we study 
the response of the pion valence momentum distribution to  the variation of (i) the effective masses of
both quark and  gluon  and (ii) the scale governing the size of the interaction quark-gluon vertex.
The  variation range of 
the three input parameters in the ladder  BSE of a $q\bar q$ bound 
system in Minkowski space, is suggested by the corresponding quantities  suggested 
by LQCD calculations
(see, e.g., Refs.~\cite{DuPRD14,Rojas2013,Oliveira:2020yac}).
 
  Our approach relies on the use of the
 ladder one-gluon exchange kernel, assuming that the effect of non-planar 
 diagrams are $N_c$ suppressed.   As
recently  shown  in a  study for bosonic bound states with 
color degrees of freedom \cite{Nogueira:2017pmj}, 
  $N_c\,=\,3$ is already large enough to reduce  the nonplanar contributions 
 in the structure 
 observables of the bound state to at most 5\%, even for strongly bound systems. 
   We solve the BS equation  adopting the technique based on (i) 
the Nakanishi integral representation of the BS amplitude and (ii) the LF projection of the BSE
(following the initial elaboration of 
Ref.~\cite{Carbonell:2010zw} and the further developments of  Refs.
 \cite{dePaula:2016oct,dePaula:2017ikc}).
The dynamical inputs in the model are   the constituent quark mass, the effective gluon mass, ranging
between $\sim$ $\Lambda_{QCD}/10$ and $2\Lambda_{QCD}$, and the size of the 
quark-gluon vertex, of the order 
  $\sim \Lambda_{QCD}$ (see, e.g.,
 Refs. \cite{Rojas2013,Oliveira:2020yac}).
 
 In our covariant model the pion state contains an infinite set 
 of LF components, that are built by a $q\bar q$ and any arbitrary 
number of effective gluons  (as needed to obtain the bound-state pole in
the four-points Green-function). In order    to better understand the
influence of   those higher-Fock states,
 we analyze (i) the decay constant, (ii) the valence probability and its spin decompositions,
(iii) the longitudinal- and transverse-momentum distributions, with a particular attention to the
end-point behavior of the first one, (iv) the distribution amplitudes and 
(v) the 3D image  on the null-plane,  using both the  LF-momentum space and 
the  3D space,
described by the covariant Ioffe-time and  the transverse 
 coordinates.
It has to be emphasized that apart from the obvious exception of the decay constant, all the
other quantities are investigated also in terms of their spin decomposition,  opening a window on 
  genuinely relativistic effects inside the pion. Such an extensive analysis 
 represents a distinctive feature of our approach
 carried out directly in Minkowski space, where the physical processes take 
 place. This is  a fundamental   step 
 toward  a future goal of 
 constructing a framework where  Euclidean and Minkowskian studies of 
 the dynamics inside hadrons can be made complementary.
 
The paper is organized as follows. In Sec.~II we outline the formalism for
solving the BS equation within 
the Nakanishi Integral Representation framework, and we briefly
illustrate   its application to the $0^-$ bound state.  In Sec.~III, 
 the  valence component of
the pion is analyzed in terms of its  spin decomposition and an analogous 
study is extended to  the LF-momentum distributions.
Section IV is devoted to the formulation of the pion decay constant in terms of the NIR, 
showing its direct relation  with the anti-aligned spin component 
of the valence wave function. In Sec. V the valence   wave function of the pion is investigated in the 3D configuration 
space associated with the null-plane, in parallel with 
the 3D representation of the pion obtained  in momentum space.
In Sec.~VI, our wide numerical exploration is presented and discussed, ranging from the pion decay
constant to the valence probability (with its spin decomposition), and from the LF-momentum
distributions to the 3D pion imaging.
 We close the work in  Sec. VII, drawing  our conclusions and presenting
 perspectives of future developments.

\section{ The Bethe-Salpeter equation and the Nakanishi integral representation}
\label{sec_NBSE}
We briefly summarize the formalism for solving the BSE
in Minkowski space for the $0^-$ quark-antiquark bound state,  within the
approach   based on both the  LF-projection technique  
 and the NIR~\cite{Nakanishi:1971}. More
 details can be attained from e.g.~Refs. \cite{dePaula:2016oct,dePaula:2017ikc},  
 (cf. the equivalent treatment within the covariant LF framework of Refs.
 \cite{Karmanov:2005nv,Carbonell:2010zw}). 
 
 For instance, in the case of a positively charged pion,  the BS amplitude 
 and its conjugate
are given in the coordinate space by (the translation invariance has been applied)
\be
\Psi(x_1,x_2,p)= e^{-ip\cdot X}~
\langle 0|T\Bigl\{ U(\tfrac x 2)~\bar D(-\tfrac x 2)\Bigr\}|\pi^+\rangle
\nonu
\bar\Psi(x_1,x_2,p)= e^{ip\cdot X}~
\langle \pi^+|T\Bigl\{ D(-\tfrac x 2)~ \bar U(\tfrac x 2)\Bigr\}|0\rangle
\label{bsa0}\ee
where (i) $U$ and $D$ are  fields with  quantum numbers corresponding to
  $u$ and $d$ quarks, respectively; (ii) $X=\eta_1 x_1+\eta_2 x_2$, with
  $\eta_1+\eta_2=1$ (in the present case $\eta_i=1/2$); (iii)  $x=x_1-x_2$ and
 (iv)  $p$ is the total 
momentum, with $M^2=p^2$   the bound-state squared 
mass. 
  It should be pointed out that 
  \be\bar\Psi(x_1,x_2,p)~\ne~ \gamma^0 
  \Psi^\dagger(x_1,x_2,p) \gamma^0~~,\ee since  one has also to fulfill the Feynman
  prescription,
  as
  encoded into the chronological operator (if one innocently applies the Dirac
  conjugation then one gets an anti-chronological ordering).
  In the momentum space,  the intrinsic components are given by
  \be
  \Phi(k,p)= \int d^4x ~e^{ik\cdot x}~\langle 0|T\Bigl\{ U(\tfrac x 2)~
  \bar D(-\tfrac x 2)\Bigr\}|\pi^+\rangle
\nonu
\bar \Phi(k,p)\int d^4x ~e^{-ik\cdot x}~\langle \pi^+|T\Bigl\{ D(\tfrac x 2)~
\bar U(-\tfrac x 2)\Bigr\}|0\rangle~.
\ee
Notice that  the chronological operator acting in the coordinate space 
generates
 the presence of 
 $+i\epsilon$ in the momentum space.
 
 The BS amplitude, $\Phi(k,p)$,  fulfills the BS equation, that in the ladder
approximation reads:
\begin{multline}
\Phi(k,p)=S(k+p/2)\int \frac{d^4k'}{(2\pi)^4} ~
S^{\mu\nu}(q)\,\Gamma_\mu(q)\\ 
\times\Phi(k',p)\,\widehat\Gamma_\nu(q)\,
S(k-p/2)\, ,
\label{bse}
\end{multline}
where (i) the  off-mass-shell constituents  have  four-momenta given by
$p_{1(2)}=p/2 \pm k$, with $p^2_{1(2)}\ne m^2$ with $m$ the constituent mass, (ii) $p=p_1+p_2$ is the total 
momentum,  (iii)  $k=(p_1-p_2)/2$ is the relative four-momentum 
and (iv) $q=k-k'$ the 
momentum transfer.  
The Dirac and gluon free propagators are given by
\be
{ S(k\pm p/2)=~i~{\psla k \pm \psla p/2+m\over  \Bigl(k\pm p/2\Bigr)^2-m^2+i\epsilon}\, ,}
\nonu
S^{\mu\nu}(q)=-i~{g^{\mu\nu}\over q^2-\mu^2+i\epsilon} \, .
\ee
Notice that the effective gluon propagator is chosen in  the Feynman
gauge.
Moreover, $\Gamma^\mu$ is the interaction vertex and $\widehat\Gamma_\nu(q)=C~\Gamma_\nu(q)~C^{-1}$ with the charge operator given by $C=i\gamma^2\gamma^0$. In  the present   model, the quark-gluon extended vertex is described by
\be
\Gamma^\mu(q)=i~ g{\mu^2-\Lambda^2\over q^2-\Lambda^2 +i\epsilon} \gamma^\mu~~,
\label{vertexff}
\ee
where $g$ is the coupling constant and  $\Lambda$  a suitable scale for 
taking  into account  the size of
the color distribution of the interaction vertex. A   quantitative estimate of
this parameter is   suggested by Refs.   \cite{Rojas2013,Oliveira:2020yac}.

\subsection{Solving the BSE for the $0^-$ bound state}
 The most general expressions for $\Phi(k,p)$  and $\bar \Phi(k,p)$ allowed by 
  the parity and the four-momenta at disposal  is  \cite{Carbonell:2010zw} (see
  also Ref. \cite{Lwel})
\be
\Phi(k,p)=  S_1 (k,p) \phi_1 (k,p)+S_2 (k,p) \phi_2 (k,p)
\nonu+S_3 (k,p) \phi_3
(k,p)+S_4 (k,p) \phi_4 (k,p)~,
\nonu \nonu
\bar \Phi(k,p)=  -S_1 (k,p) \phi_1 (k,p)+S_2 (k,p) \phi_2 (k,p)
\nonu+S_3 (k,p) \phi_3
(k,p)+S_4 (k,p) \phi_4 (k,p)~
\label{bsa}
\ee
where 
 $\phi_i$ are suitable scalar functions that depend upon: $\{k^2,\, p^2,\, k\cdot p\} $,  
 and contain the analytical behavior 
 imposed by
 the Feynman prescription, i.e.  $+i\epsilon$. They have  to fulfill  well-defined 
  properties under the exchange  $k\to -k$, according to   the anti-commutation 
  rule for the
 involved fermionic fields. 
   As a result, one has even scalar functions  for $i=1,2,4$ and an odd one
  for $i=3$,  when $k\to -k$. The allowed Dirac structures are represented
 by the $4\times 4$ matrices $S_i$,  given by 
\be
S_{1}= \gamma_5\,,\, S_{2} = {\psla p\over M}  \,\gamma_5\,,\,
 S_{3} = {k \cdot p \over M^3}  \psla p ~\gamma_5 - {1\over M} \psla k 
\gamma_5~, \nonu
S_{4} = {i \over M^2} \sigma^{\mu\nu}  p_{\mu} k_{\nu} ~\gamma_5 ~.
\label{S_structure}
\ee
The above matrices satisfy  orthogonality relations that allow one to
 reduce the BS equation (\ref{bse}) to a system of four coupled integral 
equations for $\phi_i(k,p)$.  The scalar functions can be 
 conveniently  written in terms of the NIR as follows: 
\begin{equation}
\phi_i(k,p)=\int_{-1}^1 dz'\int_0^\infty d\gamma' 
{ g_{i}(\gamma',z';\kappa^2) \over 
\left[{k}^2+z' p\cdot k -\gamma'-\kappa^2+i\epsilon\right]^3},
\label{phinak}
\end{equation}
where 
{
\be
\kappa^2 = m^2- M^2/4~~,
\label{def_kappa}
\ee and   $ g_{i}(\gamma',z';\kappa^2)$ are
called  Nakanishi weight functions (NWFs) of the scalar function $\phi_i(k,p)$.  
 Noteworthy,  $g_{i}(\gamma',z';\kappa^2)$ are real functions which are
 conjectured to be unique (cf.~the theorem on the uniqueness by Nakanishi in Ref.
 \cite{Nakanishi:1971}). Those functions  encode all the non perturbative dynamical information. 
The power of the denominator in Eq.~(\ref{phinak}) can be chosen as any 
convenient integer $\ge 3$, since we are considering the BS amplitude (see
also  Ref.~\cite{Carbonell:2010zw}).  
 The  properties of the scalar
 functions $\phi_i(k,p)$ under the exchange $k\to -k$ can
  be straightforwardly
 translated into the corresponding properties of the NWFs
 $g_{i}(\gamma',z';\kappa^2) $ under the
 exchange $z'\to -z'$, i.e.~they must be even for $i=1,2,4$ and odd for $i=3$.

As is well-known,  Eq.~\eqref{phinak} 
allows us to perform the LF projection of the BS amplitude, leading to 
 the valence component of the state (see, e.g., Refs. \cite{FSV1,FSV2} and Sec.~\ref{sec_val}).
 This motivates the application of
  the same projection to both sides of BSE, with
the aim of determining the NWFs, and eventually the full BS amplitude in Minkowski space. 
In particular, following Refs. 
\cite{dePaula:2016oct,dePaula:2017ikc}, one starts from  the coupled
system of integral equations for the scalar functions $\phi_i (k,p)$ and 
 arrives at a coupled system  for the
 NWFs, viz
\be
\label{coupls2}
\int_0^\infty d\gamma' { g_{i}(\gamma',z;\kappa^2) \over 
\left[\gamma + \gamma' + m^2 z^2 + (1-z^2)\kappa^2\right]^2} = 
iM g^2 
\nonu\times~\sum_{j}  \int_0^\infty d\gamma' \int_{-1}^1 dz'{\mathcal L}_{ij} 
(\gamma,z;\gamma', z') 
 ~g_{j}(\gamma',z';\kappa^2)\,~~, 
 \nonu\ee
where the kernel ${\mathcal L}_{ij} 
(\gamma,z;\gamma', z') $,  in the ladder approximation, can be found in full detail
 in Ref. \cite{dePaula:2017ikc}. 
 { It is worth noticing that   the
 two-scalar case  was also studied by  using 
  an ordinary linear integral equation, where on the lhs
  there is directly the NWF and on the rhs the  folding of the NWF with a 
  suitable kernel. This integral equation was obtained  exploiting 
 a uniqueness  theorem  by Nakanishi \cite{Nakanishi:1971}, assumed to be valid for the non-perturbative case and applied within the LF framework in  Ref. 
 \cite{FSV2}.  Unfortunately,   an analogous treatment for the two-fermion case is hindered by the presence of
 singularities in the interaction kernel (see below and Refs. \cite{dePaula:2016oct,dePaula:2017ikc}),  making it unclear whether or  not the Nakanishi theorem can be formally applied. Hence  
 a more careful analysis is necessary and it will be presented  elsewhere.}
  
We strongly emphasize that the kernel receives contributions from 
LF singularities originated by the treatment of the spin degrees of
freedom acting in the problem. They were successfully { taken into account} by 
{means of} the methods developed in
\cite{dePaula:2016oct} (see Ref.~\cite{YanII} for a previous
 discussion of those singularities). 
The above
set of integral equations is solved numerically by matrix manipulation algorithms, 
 after expanding the NWFs onto the Cartesian product of   Laguerre polynomials,
 for the $\gamma$ dependence, and 
  Gegenbauer ones, $C^{\lambda_i}_n$  with suitable $\lambda_i$, for the $z$
  dependence.

\subsection{ Normalization}
In order to calculate hadronic properties,  in our case the valence probability and  momentum distributions,
the BS amplitude has to be properly normalized.
 In the ladder approximation, the normalization   reads ~\cite{lurie}
\begin{multline}
Tr \Bigg[ \int {d^4 k \over (2 \pi)^4} {\partial \over \partial  p^{\prime\mu}}
 \{ S^{-1} (k-p^\prime/2) 
 \bar\Phi(k,p) \\ \times S^{-1} (k+p^\prime/2) \Phi(k,p) \} |_{ p^\prime= p}
 \Bigg] =  -i ~ 2 p_{\mu} \, .
\label{nap1b}
\end{multline}
 It is worth noting that such a normalization can be easily reverted
into the charge normalization, within the adopted ladder approximation.

By using Eq.~(\ref{bsa}) and performing the Dirac traces, the normalization condition 
turns to be:
\begin{multline}
~i\;N_c~\int {d^4 k \over (2 \pi)^4}  ~
\Bigg[\phi_1 \phi_1+\phi_2 \phi_2 +b \phi_3\phi_3+ b \phi_4\phi_4 \\ - 
4~b\phi_1 \phi_4 -4 {m\over M} \phi_2 \phi_1
\Bigg]= 1 \, , \label{normw}
\end{multline}
where $N_c$ is the number of colors and 
$
b =  \left[ (k\cdot p)^2 - k^2 M^2 \right]/M^4 \, .
$
By introducing the amplitudes $\phi_i$ given in terms of the NIR, Eq.~(\ref{phinak}), one can 
 straightforwardly perform
the analytical integration of the momentum-loop using Feynman parametrization 
in
Eq.  (\ref{normw}), 
and finally get 
\be
1= -{3 N_c\over 32\pi^2}\int_{-1}^{+1} dz' \int_{0}^{\infty} d\gamma'  \int_{-1}^{+1} dz
\int_{0}^{\infty} d\gamma 
 \nonu \times ~\int ^1_0 dv ~     v^2(1-v)^2 ~{1\over \left[
\kappa^2 + {M^2 \over 4} \lambda^2  +\gamma' v
+\gamma(1-v) \right]^3}
\nonu\times
\Biggl\{ {G(1;2)
-4{m\over M} \mathcal{G}_{21}(\gamma',z';\gamma,z) \over \left[
\kappa^2 + {M^2 \over 4} \lambda^2  +\gamma' v
+\gamma(1-v) \right]}
\nonu +{1\over 2M^2}
~\Bigl(G(3;4)
-4 \mathcal{G}_{14}(\gamma',z';\gamma,z)\Bigr) \Biggr\} ~~,
 \label{nend}
\ee
where
\be
G(1;2)=\mathcal{G}_{11}(\gamma',z';\gamma,z)+ 
\mathcal{G}_{22}(\gamma',z';\gamma,z)~~,
\nonu
G(3;4)=\mathcal{G}_{33}(\gamma',z';\gamma,z)+ 
\mathcal{G}_{44}(\gamma',z';\gamma,z)~~,
\ee
with 
$\mathcal{G}_{ij}(\gamma',z';\gamma,z)= g_{i}(\gamma' ,z' ;\kappa^2)~g_{j}(\gamma,z;\kappa^2)\, $,
and $\lambda=[v\, z'+(1-v)\, z]$.

Even in the ladder approximation, the normalization, Eq. \eqref{normw}, contains the contributions 
beyond the valence  one from the Fock expansion of the pion state,
i.e. it
takes into account 
 the infinite
sum of states with a quark-antiquark  pair and  any number of gluons (see e.g.~\cite{Sales:2001gk,Marinho:2008pe}).

We should call the attention of the reader to the fact that from the normalization
 condition
 Eq. \eqref{normw}, one can arrive at   the normalization fulfilled by
  the amplitudes of
 the pion-state Fock components (see, e.g., Refs. \cite{FSV1,FSV2}), 
 so that a probabilistic
 framework can be restored. In particular, the normalization of the 
 valence component is nothing else  than  the probability to find the component
 with the lowest number of constituents inside the pion.

\section{Valence probability and LF momentum distributions} 
\label{sec_val}
The valence probability and momentum distributions can be derived resorting to the
LF quantum-field theory methods (see, e.g., Ref.~\cite{Brodsky:1997de}), where one
defines the creation and annihilation operators for particles and antiparticles with arbitrary spin
on the null-plane, in
order to construct the generic LF Fock state. 
Actually, the valence component is defined by
 \be
 \varphi_{2}(\xi, \bm{k}_{\perp },\sigma_i;M,J^\pi,J_z) =
  (2\pi)^3 ~\sqrt{N_c}~2p^+ ~\sqrt{\xi(1-\xi)}
\nonu
 \times ~\langle 0| b(\tilde q_2, \sigma_2)
~d(\tilde q_1, \sigma_1)| 
\tilde p, M,J^\pi,J_z\rangle~,
\label{valdef}
 \ee
 where $b(\tilde q_2, \sigma_2)$  is the quark annihilation operator and $d(\tilde q_1, \sigma_1)$
 is the antiquark one, $\tilde q_1\equiv \{q^+_1=M (1-\xi),-{\bf k}_\perp\}$,  
$\tilde q_2\equiv \{q_2^+=M\xi,{\bf k}_\perp\}$ and  $\xi= 1/2 +k^+/p^+$. 
 Eq. \eqref{valdef} is related to the BS amplitude by means of  the LF projection (see the details in 
 Appendix \ref{app_val}) as follows
\be
 \varphi_{2}(\xi, \bm{k}_{\perp },\sigma_i;M,J^\pi,J_z) 
=
 {\sqrt{N_c} \over p^+} {1 \over 4}~\bar u_\alpha(\tilde q_2,\sigma_2)
 \nonu \times~
\int {dk^{-}\over 2\pi} \Bigl[\gamma^+~\Phi(k,p) \,  \gamma^+
\Bigr]_{\alpha\beta}~
  v_\beta(\tilde q_1,\sigma_1) 
\label{valbse}
~. \ee 
From Eq.~\eqref{valbse}, 
 one ultimately recognizes 
that the evaluation of the valence wave function comes from the elimination of 
the relative
LF time between the quark operators  entering  the
 BS amplitude (see also Ref. \cite{FSV1}). 

Alternatively,
the valence wave function can be obtained using the quasi-potential expansion method adapted to 
perform the LF projection of the BS equation and amplitude (see Refs.
\cite{Sales:1999ec,Sales:2001gk,Marinho:2008pe,Frederico:2010zh} for details).

As above mentioned, the Fock expansion of the interacting system allows
one to restore a probabilistic framework that the BS amplitude is not able
to  yield. In particular,   one can get    the valence probability given by  
\be
P_{val} =
{1 \over (2\pi)^3}~\sum_{\sigma_1\sigma_2}\int_{-1}^1 {dz \over (1-z^2)} 
\int
d{\bf k}_\perp
\nonu \times~
\Bigl\vert \varphi_{n=2}(\xi, \bm{k}_{\perp },\sigma_i;M,J^\pi,J_z)
\Bigr\vert^2
\label{Pval}\ee
where 
$z=1-2\xi=-2k^+/M$. As it is shown in detail in Appendix
\ref{app_val}, the valence component can be decomposed into two spin
contributions, given by the configurations we indicate as  {\em anti-aligned} and {\em
aligned}. The first configuration corresponds to a total spin of the
quark-antiquark pair $S=0$, while the second one pertains to the $S=1$
 spin-state. It is
important to emphasize that the anti-aligned configuration, yielding the largest contribution to the
pion state,  is coupled to the eigenstate of the  operator $L_z$ with
{eigenvalue  $\ell_z=0$. Within the LF framework, the $z$-component of the  orbital angular momentum is
 diagonal in the Fock space, being a kinematical generator.}
{Differently} the aligned
configuration necessarily invokes the eigenvalues $\ell_z=\pm 1$, given the
pseudoscalar nature of the pion.  
Interestingly, the presence of the 
aligned-spin contribution 
is  an unavoidable and clear 
 signature of the relativistic dynamical regime  inside the pion, since 
 the $\ell_z=\pm 1$ orbital component is related to the small
 component of the spinors in the Dirac theory. It is clear that a quantitative
 study of the aligned component has a pivotal role in understanding the
 relativistic features of the light hadrons (and the possible relativistic
 corrections for the heavier ones). 
   Finally, it should be pointed out  that the $\ell_z=\pm 1$ contribution 
  is perfectly compatible with the
 requested parity conservation (recall the minus signs 
 in the matrix $\gamma^0$ entering the representation of the parity operator).  
  
 The obtained expression for $\varphi_2$ is 
  \be\varphi_{2}(\xi, {\bf k}_\perp,\sigma_i;M,J^\pi,J_z) 
=
- {\sigma_2\over 2} ~\sqrt{N_c}~\sqrt{1-z^2}
\nonu\times~\int {dk^{-}\over 2\pi}
~\Biggl\{  \delta_{\sigma_2,-\sigma_1}~ \Bigl[ \phi_2(k,p) +
 \Bigl({k^- \over 2M}+{z\over 4}\Bigr)    ~ \phi_3(k,p)\Bigr]~ 
  \nonu { -} \delta_{\sigma_2,\sigma_1} ~
 { k_{L(R)}\sqrt{2} \over  M} ~  \phi_4(k,p) \Biggr\}=
 - {\sigma_2\over 2} ~\sqrt{N_c}~\sqrt{1-z^2}
\nonu \times~\Biggl\{  \delta_{\sigma_2,-\sigma_1}
\psi_{\uparrow\downarrow}(\gamma,z)
{  \mp  
 \delta_{\sigma_2,\sigma_1} ~e^{\mp i\theta}}~ \psi_{\uparrow\uparrow}(\gamma,z)\Biggr\}
\,,
\label{spin_cont2}\ee
where $k_{L(R)}= \pm ~{1\over \sqrt{2}}(k_x  \mp i k_y)=\pm 
\sqrt{\gamma\over
2}~ e^{\mp i \theta}$, with $\gamma=|{\bf k}_\perp|^2$.
Moreover,
\be
\psi_{\uparrow\downarrow}(\gamma,z)=
\psi_{\downarrow\uparrow}(\gamma,z)=
\psi_2(\gamma,z) +{z\over 2} \psi_3(\gamma,\xi)
\nonu+ {i\over M^3} 
~\int_0^{\infty} d\gamma'~
~{\partial g_{3}(\gamma',z;\kappa^2)/\partial z\over
  \gamma+\gamma'+z^2m^2 +(1-z^2)\kappa^2 -i \epsilon} \,,
\label{psiantiparallel}
\nonu\ee
and 
\be
\psi_{\uparrow\uparrow}(\gamma,z)= \psi_{\downarrow\downarrow}(\gamma,z)={ \sqrt{\gamma} 
\over  M} ~\psi_4(\gamma,z)
~~.
\label{psiparallel}\ee

In the above equations, $\psi_i$ are given by
\be
\psi_i(\gamma,z)=-{i\over M}
\nonu \times ~\int_0^{\infty} d\gamma' 
\frac{g_{i}(\gamma',z;\kappa^2)} {[\gamma + \gamma' + m^2 z^2 + (1-z^2)\kappa^2-i\epsilon]^2} \, .
~~.\label{psi_gz}\ee
Notice that  for the two spin configurations,  $S=$ 0 and 1, one has 
the suitable { dependence upon the angle in the transverse plane, as 
dictated by
the} 
 eigenvalue  $\ell_z$, 0 and $\pm 1$, respectively.

After inserting Eq.~\eqref{spin_cont2} into Eq.~\eqref{Pval}, one can write  the valence
probability  in terms of  the valence momentum-distribution 
density, $\mathcal{P}_{val}(\gamma,z)$, i.e.
\be
P_{val} =  \int_{-1}^{1} dz \int_{0}^{\infty} d\gamma\,
\mathcal{P}_{val}(\gamma,z) ~~,
\label{pval0}\ee
where
 \be
\mathcal{P}_{val}(\gamma,z)=\mathcal{P}_{ \uparrow\downarrow}(\gamma,z)+
\mathcal{P}_{\uparrow\uparrow }(\gamma,z)~~,
\label{pval1}
\ee
with the  anti-aligned
and aligned probability densities 
defined by  
 \be
\mathcal{P}_{\uparrow\downarrow}(\gamma,z)=
\frac{N_c}{16 \,  \pi^2} |\psi_{\uparrow\downarrow}(\gamma,z) |^2
 \, ,
 \nonu\label{pval01}
\ee
and 
\be
\mathcal{P}_{\uparrow\uparrow}(\gamma,z)=
{ N_c\over 16 \,  \pi^2 }
|\psi_{\uparrow\uparrow}(\gamma,z)|^2=
\nonu
={ N_c\over 16 \,  \pi^2 }{\gamma\over M^2}~|\psi_4(\gamma,z)|^2~~.\label{pval02}
\ee
Recall that $|k_{L(R)}~\sqrt{2}|^2=|{\bf k}_\perp|^2=\gamma$.

The valence  longitudinal  and transverse LF momentum distribution densities 
are obtained by properly integrating  the valence probability density 
$\mathcal{P}_{val}(\gamma,z)$.  The longitudinal-momentum distribution,
 with its spin decomposition, is given by 
 \be
\phi(\xi)=\phi_{\uparrow\downarrow}(\xi)+\phi_{\uparrow\uparrow }(\xi)
~~ ,
\label{LMD_val} \ee
 with
 \be
 \phi_{\uparrow\downarrow(\uparrow\uparrow )}(\xi)=
\int^\infty_{0} d\gamma \,\mathcal{P}_{\uparrow\downarrow(\uparrow\uparrow )}
(\gamma,z)~~.
\label{LMD_val11}\ee

\begin{figure}[htb]
\begin{center}
\epsfig{figure=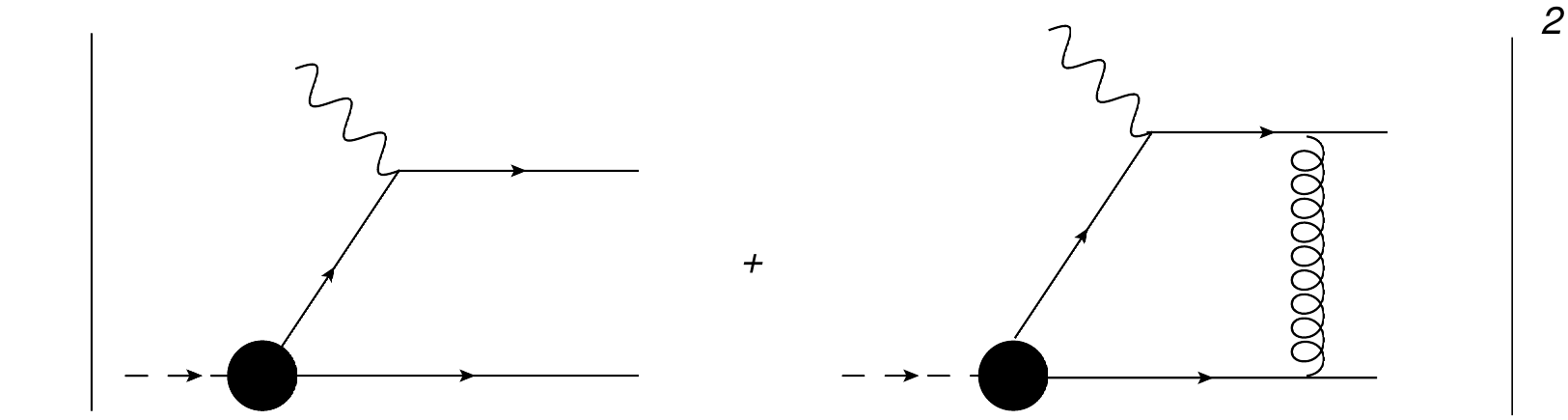,width=8cm}
\caption{Pion virtual photon absorption square amplitude. Lowest order approximation (left-frame) and 
one-gluon exchange contribution (right-frame).}\label{fig1}
 \end{center}
\end{figure}

For the transverse-momentum distribution one has
\be
P(\gamma)= P_{\uparrow\downarrow}(\gamma)+P_{\uparrow\uparrow }(\gamma)
 ~,
\label{DEFTMD0}\ee
with
\be
P_{\uparrow\downarrow(\uparrow\uparrow )}(\gamma)=
\int^1_{-1} dz \mathcal{P}_{\uparrow\downarrow(\uparrow\uparrow )}(\gamma,z) ~~.
 \label{DEFTMD}
\ee
It should be pointed out that  $\phi(\xi)$ is the unpolarized structure 
function, which one can access in the deep inelastic limit 
of the { leading order } virtual photon absorption 
process,  as  illustrated by the diagram on the left side of Fig.~\ref{fig1}. In
this case, where the final states are given by   $q\bar q$ plane waves,
 the description of the inclusive process that allows one to extract the
longitudinal distribution can be obtained by integrating on the final states, 
 so that  its pictorial representation is yielded by  
 the box diagram for  on-mass-shell quarks, as discussed in Ref. \cite{FrePRD94}
 (for a description of the deep-inelastic  scattering in an exactly solvable
 LF model with a
 confining interaction, see, e.g., Ref. \cite{Pace:1997di}). The right side of Fig.~\ref{fig1} represents the contribution from the one-gluon
 exchange in the final state, that comes from the expansion of the Wilson line, needed for assuring the 
color gauge invariance of the  quark correlator entering in the description of the deep-inelastic process
(see, e.g., Ref.~ \cite{Collins:2011zzd}). It has to be pointed out that such a  contribution in  the final state 
  is also necessary for obtaining non-vanishing T-odd transverse momentum distributions
  (TMDs) \cite{Boer:1997nt}.

\section{Decay constant}\label{piondecayconst}
 A basic observable that one has to reproduce for assessing a given  approach is 
surely the pion decay constant, $f_\pi$. It  is defined  in terms of the BS
 amplitude by  (see, e.g., Ref. \cite{Maris:1997hd} for details)
\begin{equation}
i \, p^\mu f_{\pi} = N_{c} \int {d^4 k\over (2\pi)^4} \mbox{Tr} [\, \gamma^\mu \, \gamma^5 \, \Phi(p,k) ] \, ,
\end{equation}
 Contracting with $p_\mu$  and using the decomposition of the BS amplitude given by Eq.~\eqref{bsa},
one can perform the trace and obtain
\begin{equation}\label{fpitrace}
i \, M^2 f_{\pi} = - 4 \, M \,  N_{c}\int {d^4 k\over (2\pi)^4}  \phi_{2} (k,p). 
\end{equation}
 It is worth noting that the decay constant is determined only by one 
 component (even under the exchange $1\to 2$) of the BS amplitude.
 
 By using LF variables, one can manipulate  Eq. \eqref{psi_gz} and get
 \be
 i \, M^2 f_{\pi} = - {4} ~M \,  N_{c}~{1\over 2}
 \int {d {\bf k}_\perp\over (2\pi)^2}\int {d k^+\over 2\pi}
 \psi_2(\gamma,z)=
 \nonu
 =
 - {\pi M^2\over  (2 \pi)^3} ~ \,  N_{c}~
 \int {d \gamma}\int_{-1}^1 {d z\over 2\pi}
 \psi_2(\gamma,z)
 =\nonu
 = i
 {N_c M\over 8\pi^2} \int_0^\infty d\gamma  \int_{-1}^1 dz~
\int_0^{\infty} d\gamma' 
\nonu \times
\frac{g_2(\gamma',z;\kappa^2)} {[\gamma + \gamma' + m^2 z^2 + (1-z^2)\kappa^2-i\epsilon]^2} \, .
\ee
Hence, within the NIR approach 
the final expression  for the decay constant  reads
\begin{equation}\label{fpi}
 f_{\pi} =  {N_{c}   \over 8\pi^2 \, M} \int_{-1}^1 dz'\int_0^\infty d\gamma' 
 { g_{2}(\gamma',z';\kappa^2) \over  \left[\gamma' + \kappa^2 + z'^2 \, M^2/4\right]} \,  .
\end{equation}
It should be recalled that $g_2$ is properly normalized through 
Eq. \eqref{nend}. 

Equivalently, one can proceed from Eq. \eqref{fpitrace}, by carrying out the 4D
integration on the NIR of the scalar function $\phi_2(k,p)$ (see Eq.
\eqref{phinak}) and eventually
obtaining the result in Eq. \eqref{fpi}. As a matter of fact, one has
\be
\int {d^4 k\over (2\pi)^4} { 1 \over 
\left[{k}^2+z' p\cdot k -\gamma'-\kappa^2+i\epsilon\right]^3} 
\nonu={i\pi^2 \over 2(2\pi)^4} {1\over  \left[-\gamma'-\kappa^2 -z'^2 \, M^2/4\right]} \, ,
\label{fpi_cov}\ee
where, after applying the change of variable $q=k+z'p/2$, a Euclidean
4D integration has been performed,  since the analytic structure is known. Hence, inserting in Eq. \eqref{fpitrace}
first    Eq. \eqref{phinak} and then the result given in Eq. \eqref{fpi_cov},
  Eq. \eqref{fpi} is re-obtained.

 Interestingly,  one can re-express $f_\pi$ in terms of 
the anti-aligned
component, i.e. $S=0$, given in Eq. \eqref{psiantiparallel}, once an integration
by part of the third term is performed, viz.
\begin{equation}\label{fpival}
f_{\pi}
=~ i~\frac{\pi \,N_c}{(2\pi)^3}\int_0^\infty d\gamma
 \int_{-1}^1 dz~ \psi_{\uparrow \downarrow} (\gamma,z) \, .
\end{equation}
 It is worth noticing that this expression is  expected, since the famous argument applied for 
explaining
the prevalence of the muonic channel  in the pion decay is based on both 
 the helicity conservation applied to a $S=0$ state {\em and} the phase space
 constraint.

\section{ The pion image on the null-plane} 
\label{sec_ioffe}

 The   analysis of the {deep-inelastic scattering (DIS)} is  usually  carried out 
 in  the infinite-momentum frame. In addition, one can study the DIS processes  in
 the target frame,  adopting the configuration space. Then, one 
is able to establish a   framework where a more detailed investigation of the 
space-time structure of the hadrons can be performed. In particular, a fruitful and
far-reaching 
example of the above mentioned program is the introduction of the so-called 
Ioffe-time  \cite{IoffePLB1969,Gribov,PestieuPRL1970}. In the laboratory frame,
it has a relevant role 
 in
addressing the issue of a  quantitative description of the interplay  
  between  dynamical regimes governed by short- and
long-range interactions. Indeed, one can also introduce a covariant realization (see,
e.g., Ref. \cite{Miller:2019ysh}) of the
Ioffe-time,  with the same aim of studying the relative importance of 
short and long light-like distances, e.g. probed in DIS processes. This covariant
expression is particularly useful for 
the analysis of the valence wave function, which, like  all  LF amplitudes in the Fock
expansion of the bound state, is a LF-boost invariant quantity 
and  properly
   encodes 
   information on the dynamics inside the hadron.
It should be pointed out that we use the Ioffe-time in the context  
  of the BS amplitude, i.e. the transition amplitude from
   a hadronic state to the vacuum, while, e.g., in Refs. 
   \cite{Belitsky:2005qn,Radyushkin:2017cyf} the focus is on the {\em generalized parton
   distributions} and/or the {\em transverse-momentum distributions}.

As is well-known, the  null-plane $x^+=t+z=0$ defines a  three-dimensional hypersurface,
 with space-like distances, 
where the LF wave functions  live. The configuration space 
 associated with this hypersurface
has {\em longitudinal} coordinate $x^-=t-z $ and  transverse position ${\bf x}_\perp$. 
The coordinate $x^-$ is   called ($c=1$)
 Ioffe-time \cite{IoffePLB1969,Gribov,PestieuPRL1970}, with the covariant version given by 
 $x\cdot p$ that in the target frame becomes $ x^-p^+/2$, { on the null-plane}. Notably,  in the
 target frame, when the DIS regime is reached,  the Ioffe-time measures 
 the light-like distance between the production of a $q\bar q$ pair by the virtual
 photon and its  interaction inside the hadron.
  Moreover, one can
  quickly realize that the Bj\"orken $x_{Bj}$,
  or { more precisely}, the longitudinal fraction $\xi=p^+_1/p^+$, is the   
  variable conjugated to the
  Ioffe-time, once the Fourier transform of the matrix elements of the electromagnetic current
  correlator  is analyzed for obtaining DIS structure functions. As 
    suggested by  phenomenological analyses (see,
  e.g., Refs. \cite{DelDuca:1992ru,Braun:1994jq}), one  realizes that the Ioffe-time
  (also indicated  as the  coherence length of the $q\bar q$ pair) is  
  $\propto~1/Mx_{Bj}$. This follows from  the energy-time uncertainty, involving
   the 
  $q\bar q$ pair off-shell energy  and the
  time interval between its production and interaction with the hadronic medium, 
  in the target frame. Hence, longer and longer  coherence lengths
  pertain to the  values of $x_{Bj}$, where the QCD dynamics is dominated by the IR  regime.
    This enforces the
  relevance of quantities that depends upon the Ioffe-time, when we aim at disentangling the different
    light-like distances  that the virtual photon probes, and eventually 
    shedding light on, e.g., 
   higher Fock states production, onset of the 
  confinement, valence structure, etc. 

In  Sec.~\ref{sec_val}, the valence wave function, $\varphi_2$,
 has been introduced  by considering its dependence upon the  momentum-space variables, i.e.
$\{\xi=k^+/p^+,
{\bf k}_\perp\}$.  Obviously, one can study
the valence amplitude also  in the configuration space, where the dependence results
 to be
  upon 
the  coordinates $\{\tilde z=x^- p^+/2,{\bf b}\}$~
\cite{Miller:2019ysh} 
(notice that  the LF-boost invariant definition of the Ioffe-time has been used and ${\bf b}\equiv {\bf x}_\perp$).
 
 {Rather than  the Fourier transform of 
 $\varphi_2(\xi,{\bf k}_\perp,\sigma_i;M,J^\pi,J_z)$, it is physically more
 interesting to address the  distribution probability in the coordinate space,
   in  analogy with the study  carried out in the 
 LF-momentum space, where the distribution is  given in Eq. \eqref{pval1} 
 and fulfills the sum rule
 \eqref{pval0}. In view of this, it is better
 to consider the Fourier transform of the two spin components $\psi_{\uparrow\downarrow}$ and
 $\psi_{\uparrow\uparrow}$,  Eqs.
 \eqref{psiantiparallel} and \eqref{psiparallel}, that reads
\be\label{phibgm}
\tilde\psi_{\uparrow\downarrow(\uparrow\uparrow)}(\tilde z,{\bf b})=~
{\rm e}^{-\frac{i}{2} \,\tilde z}~e^{i\ell_z \theta_{\hat {\bf b}}}\int^1_{-1}\frac{ dz}{4\pi}
~{\rm e}^{\frac{{i}}{2}z \,\tilde z}
\nonu \times \int\frac{ d{\bf k_\perp}}{(2\pi)^2}~{\rm e}^{{i}{\bf k_\perp}\cdot {\bf b}}~
e^{i\ell_z (\theta_{\hat {\bf k}}-\theta_{\hat {\bf b}})} \psi_{\uparrow\downarrow(\uparrow\uparrow)}
(|{\bf k}_\perp|^2,z),
\ee
where (i) $\ell_z=0, 1$, for $S=0,1$ respectively, 
and (ii) for   $x^+=0$ the scalar product 
$x\cdot k$ reduces to   
 $x\cdot k=x^- k^+/2- {\bf b}_\perp \cdot {\bf k}_\perp$ (in our convention). The amplitudes
 $\psi_{\uparrow\downarrow(\uparrow\uparrow)}
(|{\bf k}_\perp|^2,z)$ vanish for $z$ outside the interval  $[-1,1]$.

 Collecting the  results presented in Appendix \ref{app_OAM}, one can  
  write the Fourier transform of the two  spin components,  Eqs.
  \eqref{psiantiparallel} and \eqref{psiparallel}, 
  in terms of  auxiliary amplitudes, where the leading asymptotic behavior for large
$b$ is factorized out, i.e. 

\begin{eqnarray}\label{psitilde}
&& \tilde\psi_{\uparrow\downarrow}(\tilde z, {\bf b} )={\rm e}^{-b\,\kappa} {\rm e}^{-\frac{i}{2} \,\tilde z} \chi_{\uparrow\downarrow}(\tilde z, {b} )\, , \nonumber \\
&& \tilde \psi_{\uparrow\uparrow}(\tilde z, {\bf b} )=e^{i \theta_{\hat {\bf b}}}{\rm e}^{-b\,\kappa} {\rm e}^{-\frac{i}{2} \,\tilde z} \chi_{\uparrow\uparrow}(\tilde z, {b} )\, ,
\end{eqnarray}
 with  $\kappa$ given in  Eq. \eqref{def_kappa} and (recall that the product $z ~g_3(\gamma',z;\kappa^2)$ i s even in $z$)
\be
\chi_{\uparrow\downarrow}(\tilde z, {b} ) 
= -{{i}\over M} ~{{\rm e}^{b\,\kappa} \over 4 (2\pi)^2 } 
\int_0^\infty d\gamma' \int_{-1}^1 dz\,\cos(\tfrac {z \tilde z}{2}) 
\nonu
\times ~\Biggl\{ F'_0(z,\gamma', b) 
 ~ g_{2}(\gamma',z;\kappa^2) 
+ F'_0(z,\gamma', b\,)~
 {z\over 2}~ g_{3}(\gamma',z;\kappa^2)
\nonu
 -{1\over M^2} 
\, F_0(z,\gamma', b)   \, 
~{\partial\over \partial z } g_{3}(\gamma',z;\kappa^2) \Biggr\}\, ,
\nonu \nonu
 \chi_{\uparrow\uparrow}(\tilde z, {b} )=
-{{ i}\over M^2}~{{\rm e}^ {b\,\kappa} \over 4(2\pi)^2}\int_0^\infty d\gamma' 
\int_{-1}^1 dz ~\cos(\tfrac {z \tilde z}{2}) 
\nonu \times\, F_1(z,\gamma', b) 
 \,  g_{4}(\gamma',z;\kappa^2)\, , 
\label{psit2}
\ee
Notice that $\chi_{\uparrow\downarrow}$  and $\chi_{\uparrow\uparrow}$  
are symmetric by $\tilde z\to -\tilde z$, i.e. the inversion of the light-like axis.

We will provide results for the amplitudes $ \chi_{\uparrow\downarrow}(\tilde z, {b} )$ and 
$\chi_{\uparrow\uparrow}(\tilde z, {b} )$, where the exponential fall-off in $b$ is factorized out,
for the sake of   presentation. 

 From the above elaboration, one can obtain the  probability density  
in the 3D space $\tilde z \otimes {\bf b}_\perp$, that reads
\be
\tilde {\cal P}(\tilde z,b)= N_c ~e^{-2b\kappa}~\Bigl[ | \chi_{\uparrow
\downarrow}|^2 +| \chi_{\uparrow
\uparrow}|^2\Bigr]
\ee
and it fulfills by construction the following relation
\be
P_{val}=~\int_{-\infty}^\infty d\tilde z~\int d{\bf b}~\tilde {\cal P}(\tilde z,b)
\ee
Finally, we observe that the null-plane components in Eq.~\eqref{psitilde} at $\tilde z=0$ 
can be directly obtained  from Euclidean-space calculations, once the spin components of the transverse 
amplitude are defined as follows
\begin{equation}\label{phitildeT}
\tilde \varphi^T_{\uparrow\downarrow}(b)=\Bigl|\tilde\psi_{\uparrow\downarrow}(0, {\bf b} )\Bigr|\quad \text{and} \quad 
\tilde \varphi^T_{\uparrow\uparrow}(b)=\Bigl|\tilde\psi_{\uparrow\uparrow}(0, {\bf b} )\Bigr|\, .
\end{equation}
The above quantities come from the integration over $\frac12~dk^+dk^-$ of the BS amplitude  (leading to $x^+=x^-=0$).
Notably, 
 given   the analytic properties of the  NIR,
 an equal result can be obtained if one integrates  on $\imath ~dk^0_Edk^3$, 
 {where} $k^0_E$ is the 
Euclidean momentum component (see Ref.~\cite{Gutierrez:2016ixt} for the 
analytical details). Thus, it could be
of interest to compare  the transverse functions obtained  by  direct  calculations in the
Euclidean space for the pion BS amplitude {with} the ones evaluated by   solving the BSE in Minkowski space through the NIR.

\section{ Quantitative study }

This section is devoted to  a wide presentation  of a quantitative study of  the charged pion structure,  carried out  within the approach 
previously outlined.
In particular, we discuss  results for   (i) the   
decay constant, (ii)   the valence  probability and its spin
decomposition, (iii) the valence longitudinal- and transverse-momentum 
 distributions,  and finally (iv) the 3D image of the pion, in the space 
described by 
the Ioffe-time and the transverse coordinates, i.e. $\{\tilde z,{\bf b}\}$.

It is important to remind that the pion BSE amounts to a set of four coupled  
integral equations
 { for the scalar functions $\phi_i(k,p)$ present in Eq. 
\eqref{bsa} (see Refs.} \cite{dePaula:2016oct,dePaula:2017ikc} for
details), and it is formally 
  transformed  into a set of coupled integral equations for the four
NWFs $g_i$ (see Eq.~(\ref{coupls2})), 
 within the NIR approach. In turn, this set of equations can be expressed as a  generalized eigenvalue problem. 
 In order to 
 accomplish
this step, we exploit the  property of 
  the NWFs to be  real functions,
depending upon real variables (one compact and the other non compact) by  expanding them onto 
 a bi-orthonormal   basis (for details see \cite{dePaula:2017ikc}). 

As discussed in Sec. \ref{sec_NBSE}, while solving the BSE in ladder
approximation, one has three input parameters:  (i) 
   the   constituent-quark mass 
 and  the  gluon  effective one, indicated by $m$ and $\mu$ respectively;  
 (ii) the scale $\Lambda$ of the
extended interaction vertex. 
The  values of the adopted  parameters
cover  a fairly broad spectrum, including the values
 inspired by LQCD results. In practice, we have considered eleven
sets of values, where  $m$  ranges from $187$ to $255$ MeV,  $\mu$ from $0.15~m$
to $2.5 ~m$ (i.e. from about $30$ to $600$ MeV) and $\Lambda$ from $m$ to $2m$,
to be of the same magnitude $\Lambda_{QCD}$ as suggested in
\cite{Rojas2013,Oliveira:2020yac}.

 For completeness, we specify  the basis and truncation scheme adopted in our calculations and our numerical accuracy.
We use for each NWF  an expansion of the form
\begin{equation}
\label{Eq:g_exp}
g_i(\gamma,z;\kappa^2)=\sum_{k=0}^{N_z}\sum_{n=0}^{N_\gamma}A^i_{kn}G^{\lambda_i}_{2k+r_i}(z)\mathcal{L}_{n}(\gamma),
\end{equation}
where  $A^i_{kn}$ are the coefficients to be determined and the functions $G^{\lambda}_{n}$, and $\mathcal{L}_n$ are defined by
\begin{small}
\begin{equation}
\label{Eq:basis_func}
\begin{aligned}
G^{\lambda}_{n}(z)&=(1-z^2)^{(2\lambda-1)/4}\Gamma(\lambda)\sqrt{\frac{n!(n+\lambda)}{2^{1-2\lambda}\pi \Gamma(n+2\lambda)}}C^{\lambda}_n(z), \\
\mathcal{L}_n(\gamma)&=\sqrt{a}L_n(a\gamma)e^{-a\gamma/2}, 
\end{aligned}
\end{equation}
\end{small}
where $C^{\lambda}_n$ denotes a Gegenbauer polynomial and $L_n$ is a Laguerre polynomial. 
{ Moreover,
in order to speed up the convergence of the Gaussian integration, it has been chosen  $a=6/m^2$}.
 It should be noticed that because of the symmetry under $z\rightarrow -z$ one has 
\begin{equation}
r_i=\begin{cases} 
0 & ; ~~i=1, 2, 4, \\
1 & ; ~~i=3. \end{cases}
\end{equation}
The basis functions defined by \eqref{Eq:basis_func} obey the orthogonality relations
\begin{equation}
\begin{aligned}
&\int_{-1}^1 dz G_l^{\lambda}(z)G_n^{\lambda}(z)=\delta_{l n} \\
&\int_{0}^{\infty}d\gamma \mathcal{L}_j(\gamma)\mathcal{L}_l(\gamma)=\delta_{jl}.
\end{aligned}
\end{equation}
In our calculations, each index  $\lambda_i$ is  a half-integer, i.e. $\lambda_i=\ell_i+1/2$, 
with  
$\{\ell_1,\ell_2,\ell_3,\ell_4\}=\{2, 4, 6, 6\}$,  representing the best choice after checking the numerical 
convergence, also varying the number of polynomials in the
basis. 
For obtaining  the results  presented in this Section,
it was used up to $N_\gamma=N_z= 60$ basis functions for each $g_i(\gamma,z;\kappa^2)$. The checked accuracy in the coupling constant 
is at the shown significant digits; for the valence probability the accuracy is three significant digits;
for  $f_\pi$ the relative accuracy is better than 0.1\%;  
for $\phi_{\uparrow\downarrow (\uparrow\uparrow)}(\xi)$, 
the  point-wise accuracy for $0.95>\xi >0.05$ is about 3 significant digits and decreasing
towards the end-points, i.e. within the width of the lines shown  in the following figures.


}

\subsection{Static properties: the decay constant and the valence probability} 
 In our calculations of the pion structure, as 
 it is customary in solving BSE, 
 one assigns a value to the quark mass $m$ and using $m_\pi=140$ MeV, one gets 
 the binding $B$, defined by
 $B=2m-m_\pi$. Hence, once we  select a value for both the gluon mass $\mu$ and  the interaction-vertex scale
 $\Lambda$, we can proceed in solving the generalized eigenvalue problem
 where the 
   quark-gluon coupling 
constant, $g^2$, is the {\em  eigenvalue} (see Eq. \eqref{coupls2}, and notice
that beyond the ladder approximation one has to cope with 
a more general eigenvalue problem \cite{CK2006b,gigante2017bound}).  For the eleven sets of the three input parameters,
 as shown in detail 
in Table \ref{table1}, we have evaluated the following main quantities:
 (i) the adimensional coupling 
\be\alpha={g^2\over 4\pi}~(1-\mu^2/\Lambda)^2\label{alpha_s}~~, \ee
that  combines
 the bare
coupling $g^2/4\pi$ and  the factors from the two extended interaction vertexes (see
the definition  in Eq. \eqref{vertexff}), (ii) the valence probability, and its spin decompositions
 and 
(iii) the charged-pion
 decay
constant. 

In Table   \ref{table1} we also inserted   two other quantities, useful for sharpening our physical insights.
The first one is the effective strength $\bar\alpha$, given by
\begin{equation} \label{alphabar}
\bar\alpha=\frac{\alpha}{\frac{\mu^2}{m^2}+0.2} \,  ,
\end{equation}
where the value of the average  transverse-momentum  $\sqrt{\langle k^2_\perp\rangle}\sim\sqrt{0.2}\,m$  
has a close correspondence  to the characteristic scale of  the decreasing behavior shown by the 
transverse-momentum distribution in the model,  as it will
be clear when presenting results for this quantity in the next subsection. 
Loosely speaking, the denominator $\mu^2 + \langle k^2_\perp\rangle$  plays the
role of  an effective mass carried by the gluon in 
the interaction region relevant for   binding the $q\bar q$ system. The second quantity is 
 the adimensional ratio $f_\pi/m$, that   yields an estimate of the strength of the effective quark-pion coupling 
 introduced in low-energy effective approaches, like the Nambu-Jona-Lasinio one (see, e.g., Ref. \cite{RuizArriola:2002bp}).
 
 Table \ref{table1} is organized according
to increasing values of the valence probability, corresponding also to decreasing values of the effective coupling 
$\bar\alpha$,   since  both of them point to the same physical mechanism, as discussed below.
 
First of all,  after slightly tuning the parameters in the range suggested also by LQCD (see the
eighth set), one is able to  reproduce 
the experimental value of the pion decay constant, i.e. 
$f^{\rm{exp}}_{\pi^\pm}=$130.50(1)(3)(13)~MeV~\cite{PDG_2018}, as well as 
  the LQCD 
 average  value $f^{\rm{LQCD}}_{\pi^\pm}=$ $130.2~(0.8)$ MeV, as given in Ref. \cite{FLAG}. By varying the 
  sets
 of parameters,  the valence probability $P_{val} $, and  $f_\pi$,
  run  in the interval  $[0.64 - 0.71]$ and $[ 77 - 134]$ MeV, respectively. 
\begin{table*}[htp] 
\begin{center}
\caption{\label{table1} Static properties of the pion ($m_\pi=140$ MeV) for  several sets of the  input 
parameters:  (i)
the quark
mass $m$, (ii) the gluon mass $\mu/m$ (per mass unit) and (iii) the extended-vertex scale
$\Lambda/m$ (see text and  Eq. \eqref{coupls2}). The following quantities are listed: (i) the coupling constant
$\alpha$  
and the effective strength, $\bar{\alpha}_S$, defined in Eq.~\eqref{alphabar}; (ii) the valence probability,  $P_{val}$, and its decomposition in 
$S=0$ component, $P_{\uparrow\downarrow}$, and $S=1$ component, $P_{\uparrow\uparrow}$; (iii)  the charged pion decay constant, $f_\pi$, and    the  adimensional ratio, $f_\pi/m$, proportional to the quark-pion coupling in the effective approaches, at low energy. 
}
\begin{tabular}{|c|c|c|c|c|c|c|c|c|c|c|c|}
\hline
Set &$m$ (MeV)  & $ B/m$ & $\mu /m$  &  $\Lambda /m$ & $\alpha$ $\left(\bar\alpha\right)$ & $P_{val}$ & 
$P_{\uparrow\downarrow}$ & 
$P_{\uparrow\uparrow}$   & $f_\pi$ (MeV)&${f_\pi}/{m}$ \\
\hline
I & 187 &1.25  & 0.15 &  2 & 5.146 (23.13) & 0.64 & 0.55 & 0.09   & 77& 0.412 \\
II  & 255 &  1.45 &1.5 &   1 & 52.78 (21.54) & 0.65  & 0.55 &  0.10  &  112 &  0.439    \\
III & 255 & 1.45  &  2  &    1 & 78.01 (18.57)&   0.66 & 0.56 &  0.11     &  117 & 0.459  \\
IV  & 215 & 1.35   & 2    & 1 & 76.28  (18.16) & 0.67  & 0.57 &  0.11    & 98 &  0.456  \\
V    & 187 & 1.25  &  2  &    1 & 74.26 (17.68)&  0.67 & 0.56 &  0.11     &  84 & 0.449  \\
VI&255& 1.45 & 2.5 & 1& 108.87 (16.87)  & 0.68  & 0.56 & 0.11    & 122 &  0.478 \\
VII & 255 & 1.45 & 2.5 & 1.1 & 87.66 (13.59) & 0.69 & 0.56 & 0.12  &  127& 0.498  \\
VIII & 255 & 1.45 & 2.5 & 1.2 & 72.32 (11.21)  & 0.70  & 0.57  & 0.13   & 130 &  0.510 \\
IX    & 255 & 1.45  &  1  &    2  & 10.40 (8.665)&  0.70 & 0.57 &  0.14     &  134 & 0.525  \\
X &  215 & 1.35 &  1 &   2 & 10.20  (8.50) &  0.71  & 0.57  & 0.14    & 112 & 0.520  \\
XI & 187 & 1.25 & 1 & 2 & 9.96 (8.30) & 0.71 & 0.58 & 0.14  & 96  & 0.513 \\
\hline
\end{tabular}
\end{center}
\end{table*}

The valence probabilities for the anti-aligned and  aligned    spin configurations  are also 
shown in Table~\ref{table1},
 where  one recognizes the expected   prevalence of the spin $S=0$ component, ranging  from 0.55 to 0.58, and the minor role of the 
 $S=1$ configuration,
  from 0.09 to 0.14. In any case, it is worth noticing that  the $S=1$ contribution, which is exclusively relativistic in nature, 
 is by no means negligible, since the relative weight increases up to 30\%.  
 From the  Fock-expansion standpoint,  the size of such a 
 relative  weight indicates that the  
higher-Fock states are quite relevant in the description of  the pion state on the null-plane. In fact, the ladder kernel of the 
BSE when projected onto the LF hyperplane
\cite{Sales:1999ec,Sales:2001gk,Marinho:2008pe,Frederico:2010zh} takes into account an infinite number of
Fock-components beyond the valence state, built as a $q\bar q $ pair and any number of gluons. 

{The remaining} probability, $1-P_{val}$, is distributed among 
the first Fock components beyond the
valence one, {and one should} notice
the 
 consistent picture   that emerges from observing  the  correlation between the  decreasing values of  the valence probability 
and the  increasing  values of $\bar \alpha$.
This behavior is rather natural to be expected, 
 as $\bar \alpha$ 
 weights in an effective way the coupling to 
the higher Fock states present in the dynamical model.  Consequently, the larger  $\bar \alpha$, the  smaller  $P_{val}$, since more gluons 
can
 be present in the intermediate states, and the valence state becomes less likely.
As to the pion decay constant, while $f_\pi$ does not  show a regular   behavior    when $\bar \alpha$ is decreasing,
 the ratio $f_\pi/m$  has  less pronounced  variations, since 
   this ratio
combines the effect of the higher Fock states through two different quantities. 
 On one side, $f_\pi$ is  associated with the  anti-aligned component of the valence wave 
function at the origin, i.e. $\tilde z=b_x=b_y=0$ (see Eq. \eqref{fpival}), and its increasing  
indicates   a major role of  more compact configurations,
necessarily related to the higher Fock states. On the other side, the quark mass determines the binding energy,  and  larger values of $B=2m-m_\pi$ are related to 
 smaller   size of the pion, i.e.   more compact configurations take place (cf. the values of $m$ and $\bar
 \alpha$
for the sets III, IV and V, as well as IX, X and XI, where $\mu$ and 
$\Lambda$ do not vary).
 Hence, the values of  $f_\pi$ show a dependence upon   quark mass, since both quantities are driven by the relevance of  states beyond the valence one. 
 
{In conclusion, }  Table~\ref{table1} {provides two
interesting insights},
 which can be { briefly
highlighted}: 
(i) the decay constant   is influenced by the compact configurations and the pion size, thus  UV  and IR properties 
are reflected in{its actual value}, 
with the conspicuous relation to the constituent quark mass, and (ii)  {\ also the valence probability
 encodes signatures of the   
UV and IR regimes}. 
  { This  will become more} clear once the behavior
 at  the end-points of the
 longitudinal distribution and  the high-momentum tail of the transverse 
 momentum distribution are analyzed, {  and  recalling  that $f_\pi$ and $P_{val}$ are obtained from both
  distributions after properly integrating.}.

\begin{figure}[tb]
\begin{center}
\epsfig{figure=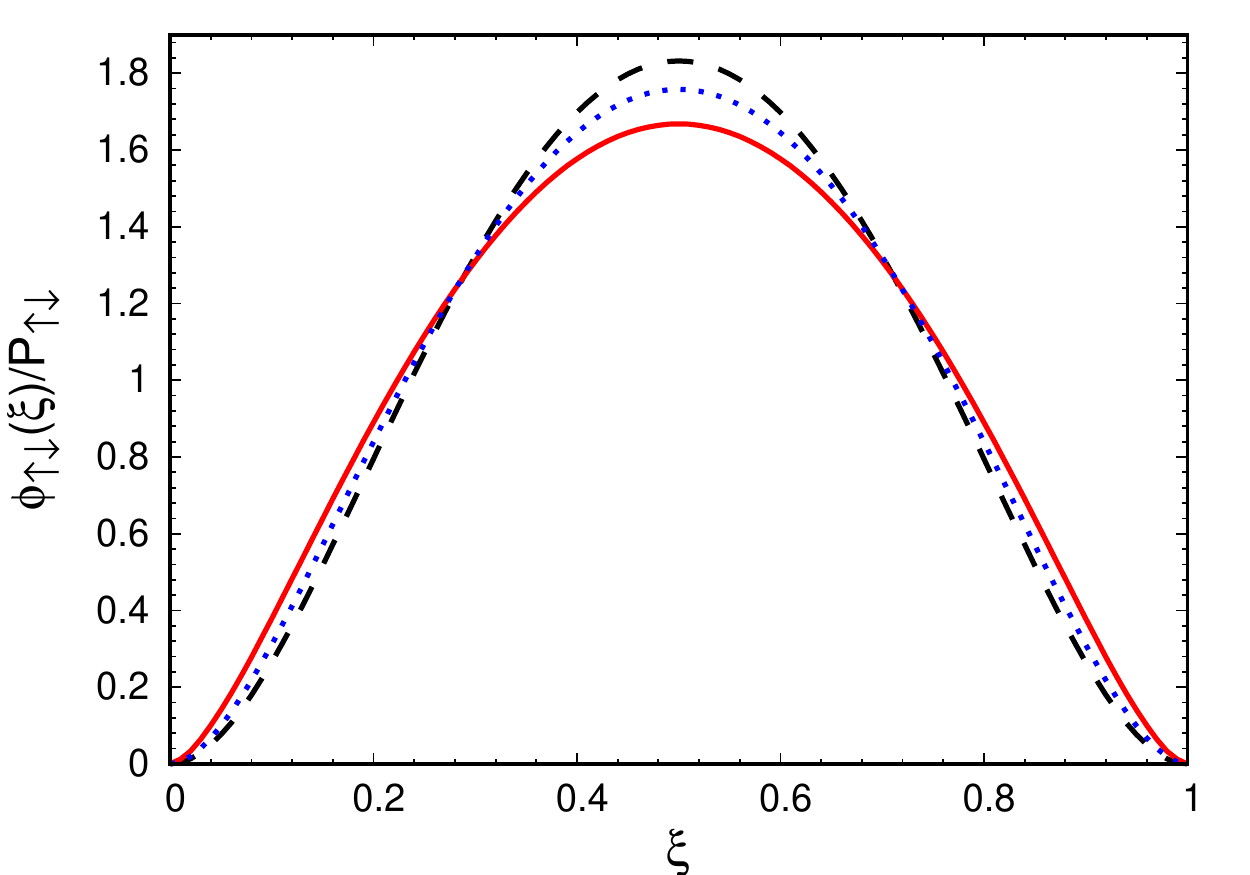,width=7.2cm}
\epsfig{figure=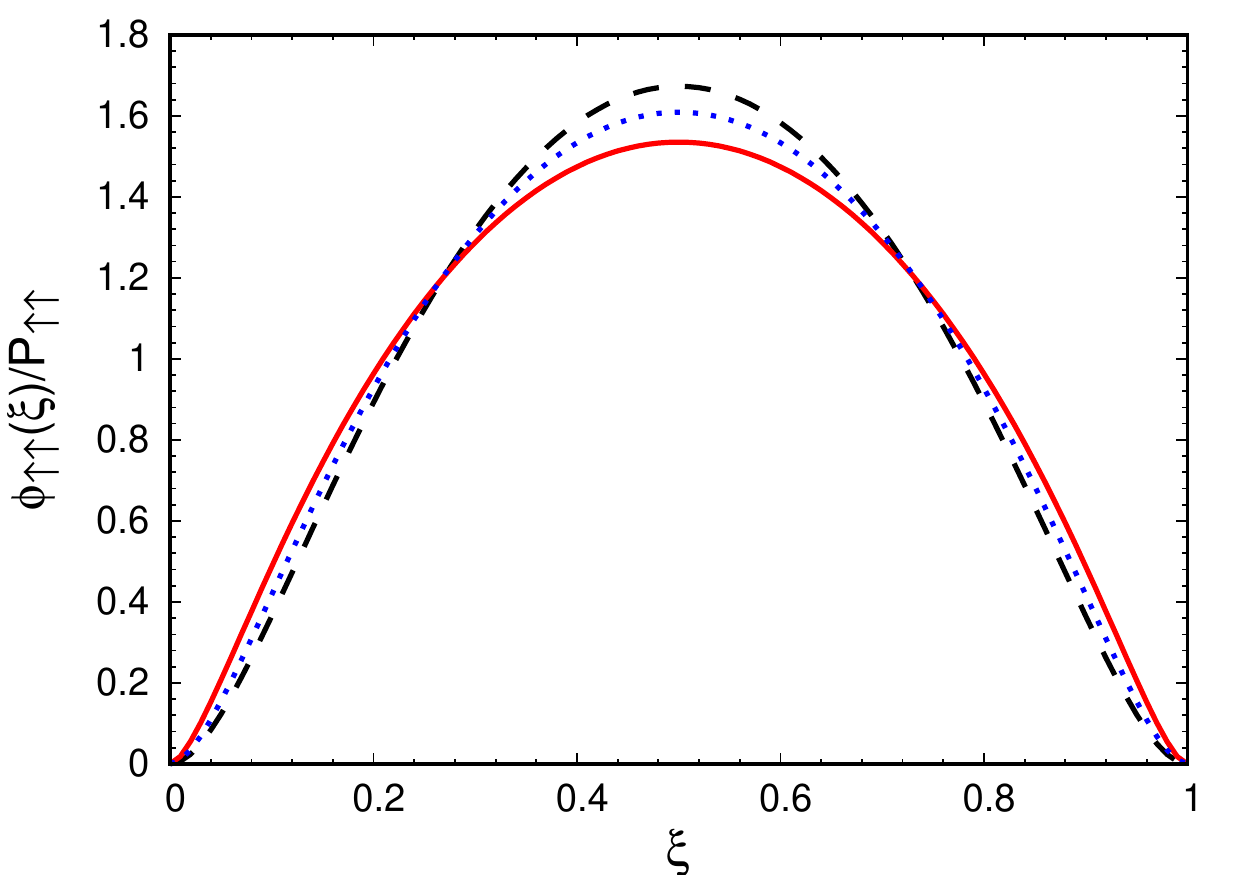,width=7.2cm}
\epsfig{figure=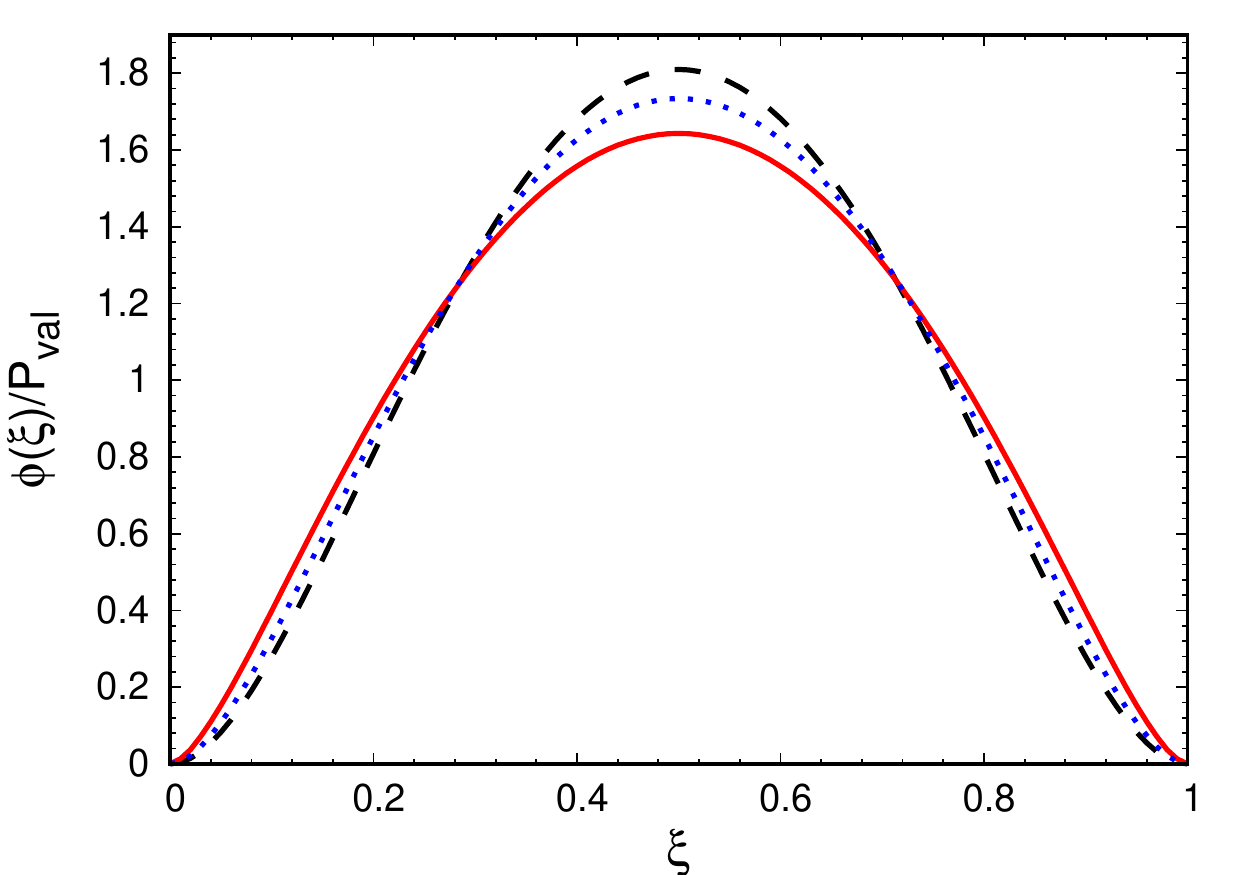,width=7.2cm}
\caption{\label{fig2}  (color online) From top to bottom:  anti-aligned, 
aligned,  and total  pion valence longitudinal-momentum 
distributions, normalized to 1, for three different sets of  
inputs parameters (cf.  Table~\ref{table1}). Dashed line: set I. Dotted line: set IV. Solid line: set VIII, yielding the experimental value for $f_\pi$.}
 \end{center}
\end{figure}

\subsection{The valence LF-momentum distributions}

In this subsection, we present the dynamical quantities predicted  for the pion on the
null-plane:
(i)    the longitudinal-momentum distribution, $\phi(\xi)$, Eq.~\eqref{LMD_val}, and
    the transverse one, ${\cal P}(\gamma)$, Eq.~\eqref{DEFTMD0}, (ii) the  valence 
    LF-momentum density, ${\cal P}(\gamma,z)$, Eqs.~\eqref{pval1},  \eqref{pval01}
    and  \eqref{pval02}, (iii) the distribution amplitude, with its spin 
    decomposition, and {its transverse counterpart}. { The calculation of
    such quantities represents the needed first step, followed by the
    suitable evolution, for  meaningful comparisons with
    experimental results, e.g.,  obtained  in  DIS and semi-inclusive DIS
    processes, like the pion parton distribution function and the transverse-momentum distributions, 
    as well as    extracted from the deeply-virtual Compton scattering, like
    the generalized parton distributions.}
    
   Let us first consider the longitudinal-momentum fraction distribution  for the valence LF
   component,  as a function of the quark longitudinal  fraction $\xi$,
 and its anti-aligned and aligned components.  It has to be stressed that 
 on the
 null-plane, the longitudinal-momentum fraction distribution  has 
  the proper support,
 and therefore it fulfills
 both normalization and momentum sum rule when we take into account the whole set of the  Fock 
 components. This is a remarkable benefit for correctly
 analyzing  DIS
 processes.
 
 The calculations are shown in
Fig.~\ref{fig2}, for some   illustrative cases, with the normalization equal to 1 (i.e. 
each distribution is divided  by the respective probability) and 
 the model  parameters  given in Table~\ref{table1}. 
 Recall that  the values for the gluon mass varies between 28 MeV $(\mu/m=0.15)$ and  638 MeV $(\mu/m=2.5)$, while 
the scale  $\Lambda$ {ranges} between  255 MeV ($\Lambda/m=1$) to 430 MeV ($\Lambda/m=2$),  covering a quite 
large {interval} of
possibilities.

Figure~\ref{fig2} shows that the decrease of the
 effective dimensionless strength of the kernel, $\bar \alpha$,   broadens $\phi(\xi)$ with a regular pattern, which is
 not too  sensitive to  
 the wide variations of both gluon mass and vertex parameter.  Notice that even in  presence of  a very light gluon, the valence longitudinal distribution  does not
show dramatic changes, although they are visible. As expected, given its genuinely relativistic nature, the aligned
distribution has a wider shape than the anti-aligned one, namely larger values of the quark momentum prevail, but the overall
impact on the total distribution is mitigated by the smaller probability associated with the $S=1$ configuration. This
observation is quantitatively corroborated by the analysis summarized in  Table \ref{table2}, where
it is    shown, for a few examples, the correlation between 
  $\bar \alpha$ 
 and the 
 exponent of the
function $(1-\xi)^\eta$, which we use,  as customary, for fitting the valence longitudinal-momentum distributions close to the 
end-points.   

Notice that the distributions are symmetric at the initial scale, while after evolving they cumulate at values $\xi\le 0.5$ 
(as it will be discussed elsewhere).
 As pointed out in the previous subsection, 
the decreasing values of the  valence probability and the corresponding increasing value of 
$\bar \alpha$ 
were put in relation with the  increasing role of compact configurations, generated by higher 
Fock states.  Plainly,  also the increasing values of the exponent $\eta$ in the fitting functions can be explained through an analogous 
 mechanism.  In fact,
the growth of the powers, shown in  Table~\ref{table2},   depletes  the  valence quark 
distribution  in the UV region, { corresponding to $\xi\to 1$.  The function  $(1-\xi)^\eta$  becomes smaller
and smaller for increasing values of $\eta$, and  { the probability of large longitudinal momenta
 decreases.} This implies  } a suppression of 
configurations with the valence $q\bar q$ pair  close together. 
 
\begin{table}[thb] 
\begin{center}
\caption{ \label{table2}  An excerpt from the set of exponents  of the fitting
  function  $(1-\xi)^{\eta}$ for $\xi\to 1$,  while varying the set of input
  parameters.  The three columns, labeled by 
$\eta_{\uparrow\downarrow}$, $\eta_{\uparrow\uparrow}$ and $\eta$,  correspond to  the anti-aligned, aligned  and total 
valence longitudinal-momentum distributions (cf. Fig. \ref{fig2}).
}
\begin{tabular}{|c|c|c|c|c|c|}
\hline
Set & ${f_\pi}/m$ & $\bar \alpha$ &$\eta_{\uparrow\downarrow}$ & 
$\eta_{\uparrow\uparrow}$ & $\eta$ \\
\hline
I & 0.412 & 23.13 & 1.81 &  1.61 & 1.77 \\
II & 0.439 & 21.54 & 1.71 &  1.50 & 1.66  \\
III  & 0.459 & 18.57 & 1.66 & 1.47 & 1.62  \\
IV  & 0.478 & 18.16 & 1.61 & 1.42 & 1.57 \\
VIII  & 0.510 & 11.21 &1.44 & 1.26 & 1.40 \\ 
IX & 0.525  & 8.665 &1.45& 1.28 & 1.40 \\
\hline
\end{tabular}
\end{center}
\end{table}%

\begin{table}[htp] 
\begin{center}
\caption{ \label{table3}  An excerpt from the set  of Mellin moments of the valence longitudinal momentum distribution, up to the sixth order.
 The  contributions of the two spin   configurations are also shown. N.B. The adopted 
  normalization for the valence longitudinal-momentum distribution is equal to $P_{val}$.
   }
\begin{tabular}{|c|c|c|c|c|c|c|}
\hline
Set & $\langle x\rangle$ &$\langle x^2\rangle$ & $\langle x^3\rangle$ & $\langle x^4\rangle$ & $\langle x^5\rangle$ & $\langle x^6\rangle$ 
 \\
 \hline
I &  0.31 & 0.18 &	  0.11 &  0.076	& 0.054	& 0.039
\\
${\uparrow\downarrow}$ & 0.27 &  0.16	&  0.10 & 0.066	& 0.047	& 0.034
\\
${\uparrow\uparrow}$  & 0.045&  0.026 &   0.017 &  0.012 & 0.008 & 0.006
\\
\hline
II   &  0.32 &  0.19	&   0.12	&  0.080	& 0.057	& 0.042
\\
${\uparrow\downarrow}$ &  0.28 &  0.16	&  0.10 & 0.068 &	 0.049	& 0.036
\\
${\uparrow\uparrow}$  &    0.048 & 0.028 &  0.018 & 0.013 &  0.009 & 0.007
\\
\hline
III  &  0.33 & 0.19 & 0.12 & 0.083 & 0.059 & 0.044
\\
${\uparrow\downarrow}$ &   0.28 & 0.16 & 0.10 & 0.070 & 0.050 & 0.037
\\
${\uparrow\uparrow}$  &    0.053 &	0.031&	0.020&	0.014&	0.011&	0.008
\\
\hline
IV  &  0.33  & 0.19	&  0.12 &	 0.082	&  0.059	& 0.043
\\
${\uparrow\downarrow}$ &   0.28 &  0.16 & 0.10 & 0.070&  0.050 & 0.037
\\
${\uparrow\uparrow}$  &    0.053 & 0.031&   0.020&  0.014  & 0.010	& 0.008
\\
\hline
VIII & 0.35 & 0.20 & 0.13 & 0.091 & 0.066 & 0.049 \\
${\uparrow\downarrow}$  & 0.28 & 0.17 & 0.11 & 0.074 & 0.054 & 0.040\\
${\uparrow\uparrow}$ & 0.068 & 0.041 & 0.027 & 0.019 & 0.014 & 0.011 \\
\hline
IX & 0.35 & 0.20 & 0.13 & 0.091 & 0.066 & 0.049 \\
${\uparrow\downarrow}$  & 0.28 & 0.17 & 0.11 & 0.074 & 0.054 & 0.040\\
${\uparrow\uparrow}$ & 0.068 & 0.041 & 0.027 & 0.019 & 0.014 & 0.011 \\
\hline
\end{tabular}
\end{center}
\end{table}

As already  noticed,  the aligned longitudinal-momentum distribution is broader than the anti-aligned one (cf.
  Fig. \ref{fig2}). This corresponds to  a softer end-point behavior,  with exponents systematically smaller than the 
  anti-aligned
   ones,  about
 $\eta^{\uparrow\downarrow}-\eta^{\uparrow\uparrow}\sim 0.2$, as shown  in Table~\ref{table2}. 
 
 The results for the Mellin moments of the valence longitudinal-momentum distributions
  are shown in Table~\ref{table3}, also with the decomposition in  spin contributions. It should be recalled that the present approach
   predicts a quite robust longitudinal fraction  carried by the valence $q\bar q$ pair, according to $P_{val}$ that ranges 
 from  60\% to 70\%. This can be assessed from the first Mellin moment, after noticing 
   that if the valence $\phi(\xi)$ is normalized to 1, and not to $P_{val}$, the value of $\langle x\rangle$ is always 0.5, due to the symmetry of the valence
  wave function
   around $\xi=0.5$. It is interesting to observe that  the moments do not change too much for the set of input parameters we have chosen, 
  and the increasing values correspond to the increasing relevance of the momentum distribution close to the end-points.
   {Such a behavior is clearly}  more evident for Mellin moments associated to  higher powers. 
   {To carry  out a detailed comparison with LQCD results and the 
   analogous outcomes from the continuum QCD approach (see, e.g. Ref.
    \cite{CloPPNP14}), one has to determine first the initial scale of  
    the valence longitudinal-momentum distribution and then  apply 
    the evolution (see, e.g., Ref. \cite{dePaula:2021}).}

The second dynamical quantity
 we have analyzed is the transverse-momentum distribution, $P(\gamma)$ (Eq.~\eqref{DEFTMD0}),  
and its decomposition in spin configurations. 
 The results, with normalization equal to $1$,  
are  presented in Fig.~\ref{fig3}  for the  same set of parameters adopted for the longitudinal-momentum distribution  in Fig.~\ref{fig2}. 
 The same general behavior found for the longitudinal-momentum distribution can be also recovered for the transverse-momentum one, namely, the size of the
 tail at large values
 of $\gamma$ becomes smaller for larger  values of $\bar \alpha$, and vice-versa. As we learned,
  this is correlated to the role played by the typical size of the spatial correlation 
 of the valence
 $q\bar q$ pair.

{The anti-aligned and aligned distributions are also shown  in Fig.~\ref{fig3}}. It is worth noticing that  
the {last one}  has a slower decay 
compared 
to the anti-aligned {case}, since it  carries a factor of $\gamma$, stemming from 
 the relativistic spin-orbit coupling, which  eventually adds a factor $\vec k_\perp$  to the momentum dependence of the 
 aligned component in 
 the valence wave function.
 The typical momentum scale that determines the  effective strength, $\bar\alpha$, i.e. $\gamma/m^2\sim 0.2$, 
can be inferred  from the behavior
of  $P(\gamma)/P(0)$ close to  $\gamma=0$, as shown in Fig. \ref{fig3}.  More explicitly,  
such value of $\gamma$   roughly indicates  the region where $P(\gamma)$ is relevant  for obtaining the actual
value of  the valence probability. {Hence, the above mentioned range of $\gamma$ 
can be associated 
with the size of the region where the interaction is 
effective in building the $q\bar q$ bound state (in momentum space). In other words, such a low-energy or IR scale  should govern the kernel  of the BSE, so that 
  it is effective in giving the
 strength necessary to create the strongly bound $q\bar q$ system, resulting  in the  pion. 
 } 

{ Summarizing the first part of the analysis, one should point out that the end-point behavior of $\phi(\xi)$ is
strongly correlated to the UV properties of the adopted kernel in the BSE, while  the range of $\gamma$, where 
$P(\gamma)$ is  large,  is governed by the size of the bound state, i.e. by the IR behavior of the interaction.}

An overall view of the pion in the 3D LF-momentum space can be obtained from  Fig.~\ref{fig4}, 
where  the LF-momentum density ${\cal P}(\gamma,\xi)$ (cf. Eqs. \eqref{pval1},
 \eqref{pval01} and \eqref{pval02}), is shown.
One should recall that the longitudinal and transverse distributions are obtained
 through the suitable integration of the density ${\cal P}(\gamma,\xi)$.
 \begin{figure}[htb]
\begin{center}
\epsfig{figure=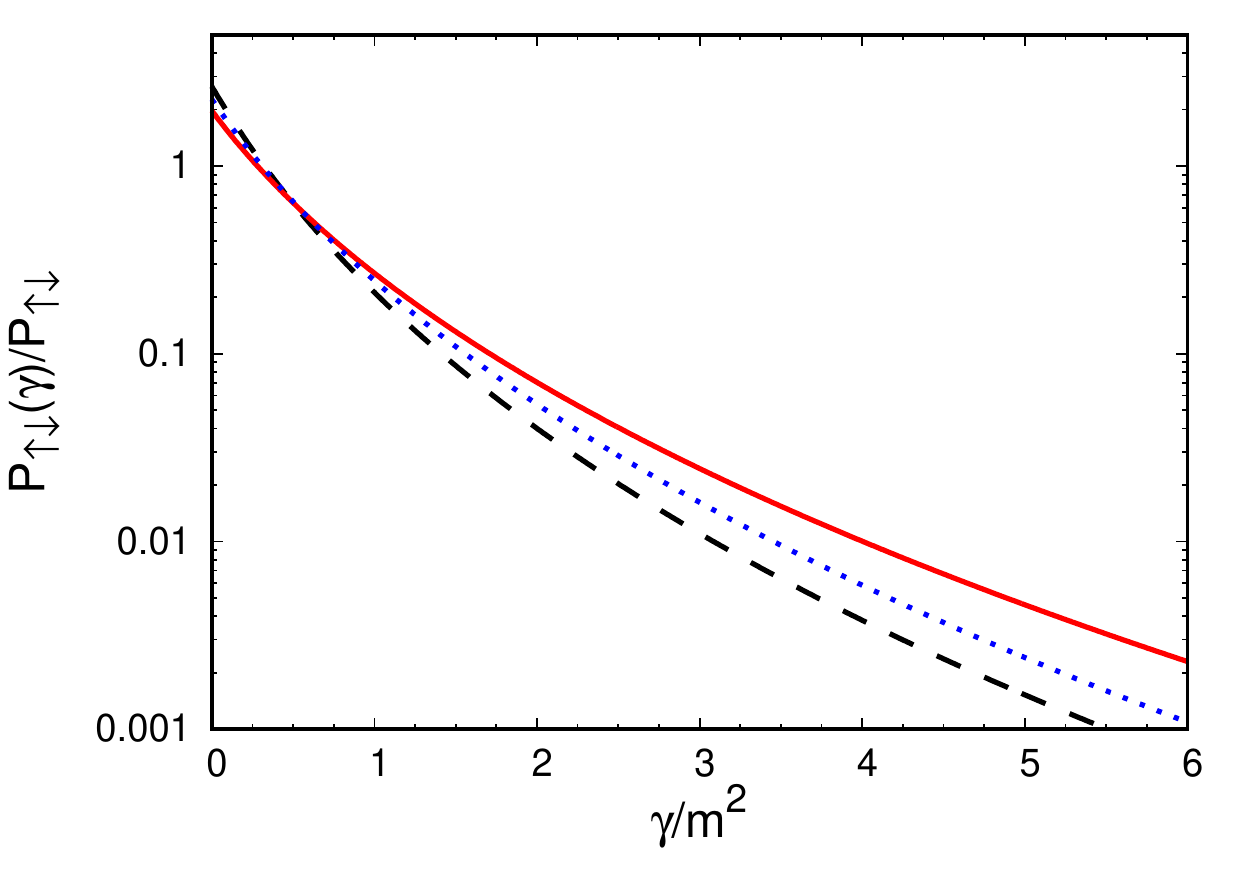,width=7.2cm}
\epsfig{figure=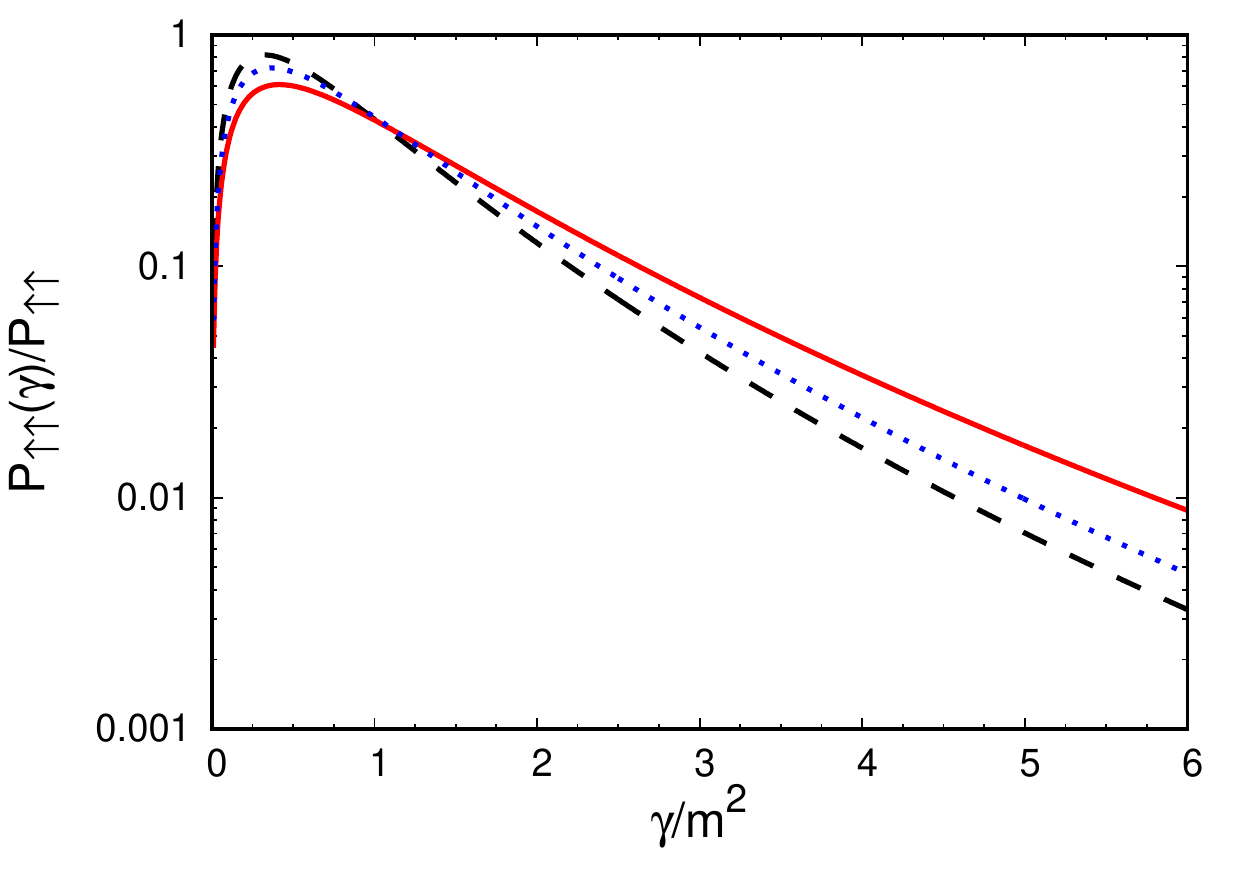,width=7.2cm}
\epsfig{figure=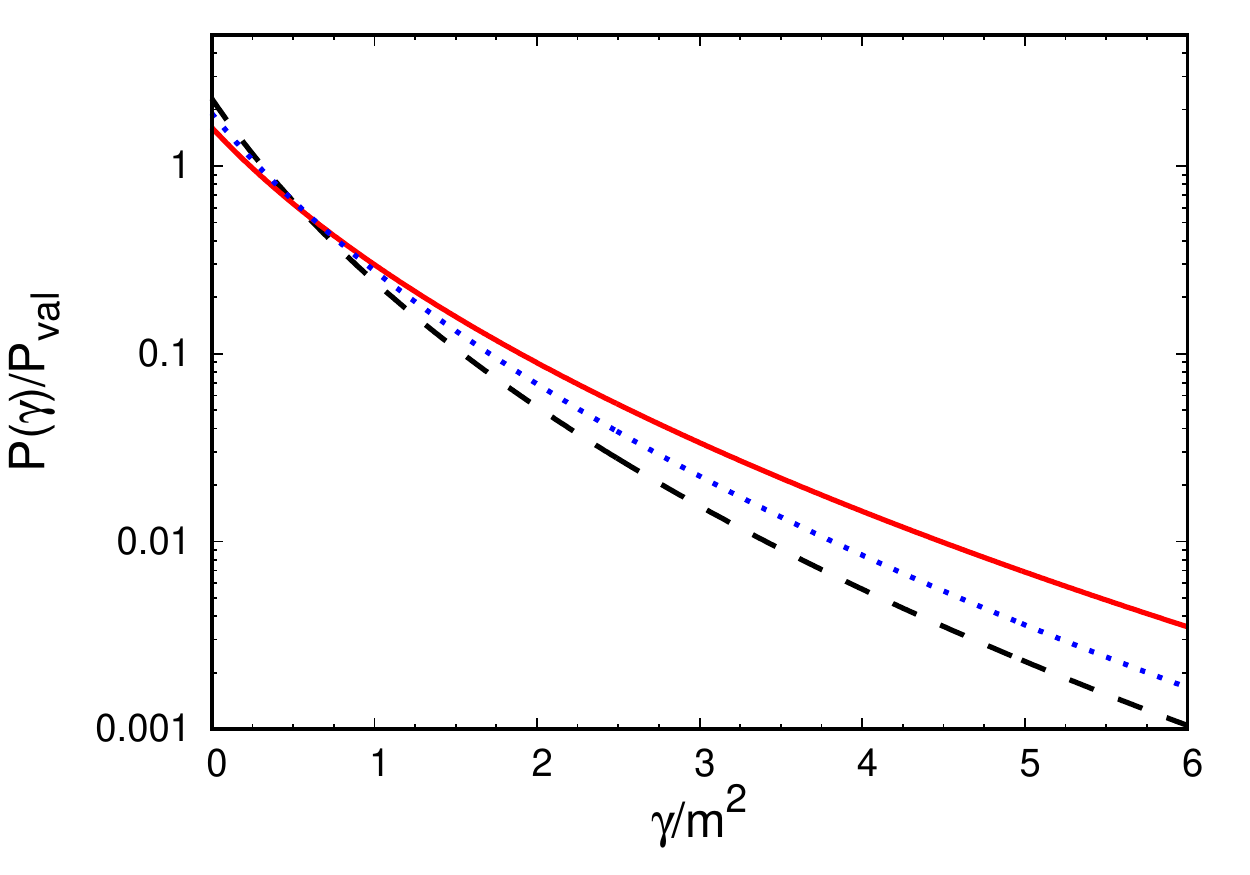,width=7.2cm}
\caption{\label{fig3} (color online) From top to bottom: 
anti-aligned, aligned  and total  pion valence transverse-momentum 
distributions, ${\cal P}(\gamma)$, normalized to 1, for three different 
sets of  input parameters  
 (cf. Table~\ref{table1}). Dashed line: set I. Dotted line:  set IV. Solid line: set VIII, yielding the experimental value for $f_\pi$.
 }
 \end{center}
\end{figure}

\begin{figure}[htb]
\begin{center}
\epsfig{figure=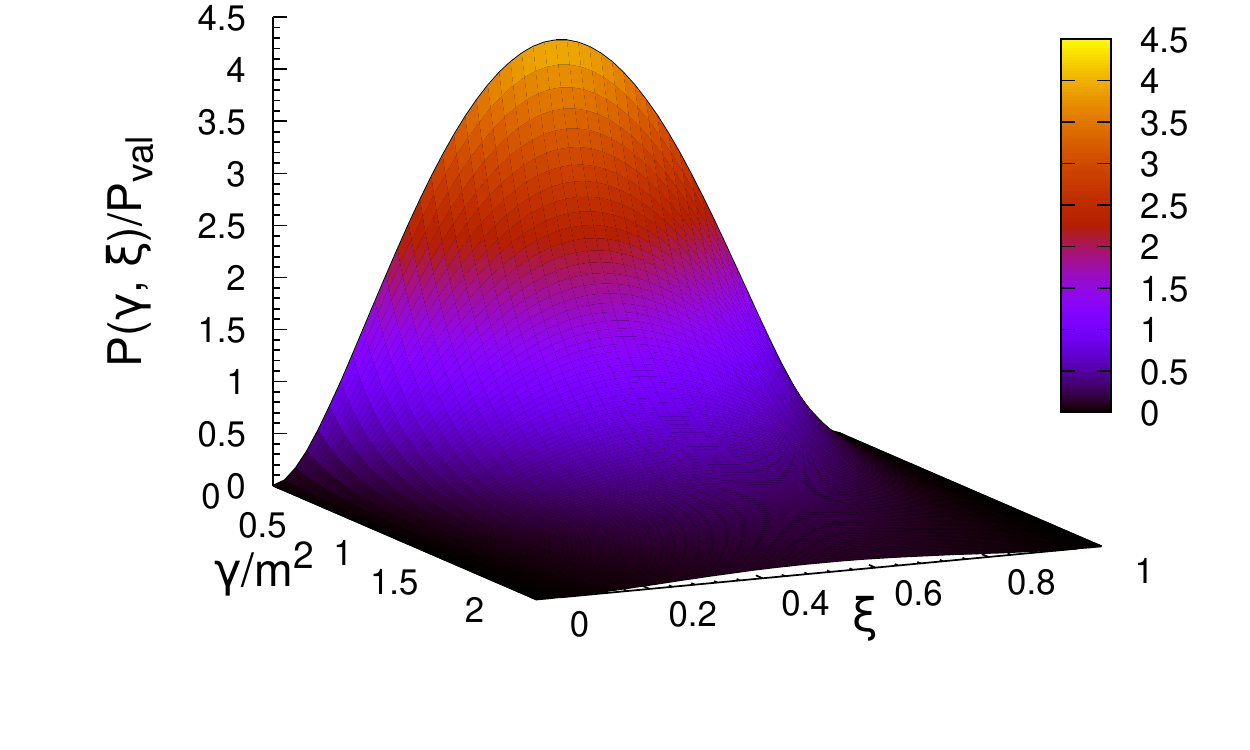,width=8cm}
\epsfig{figure=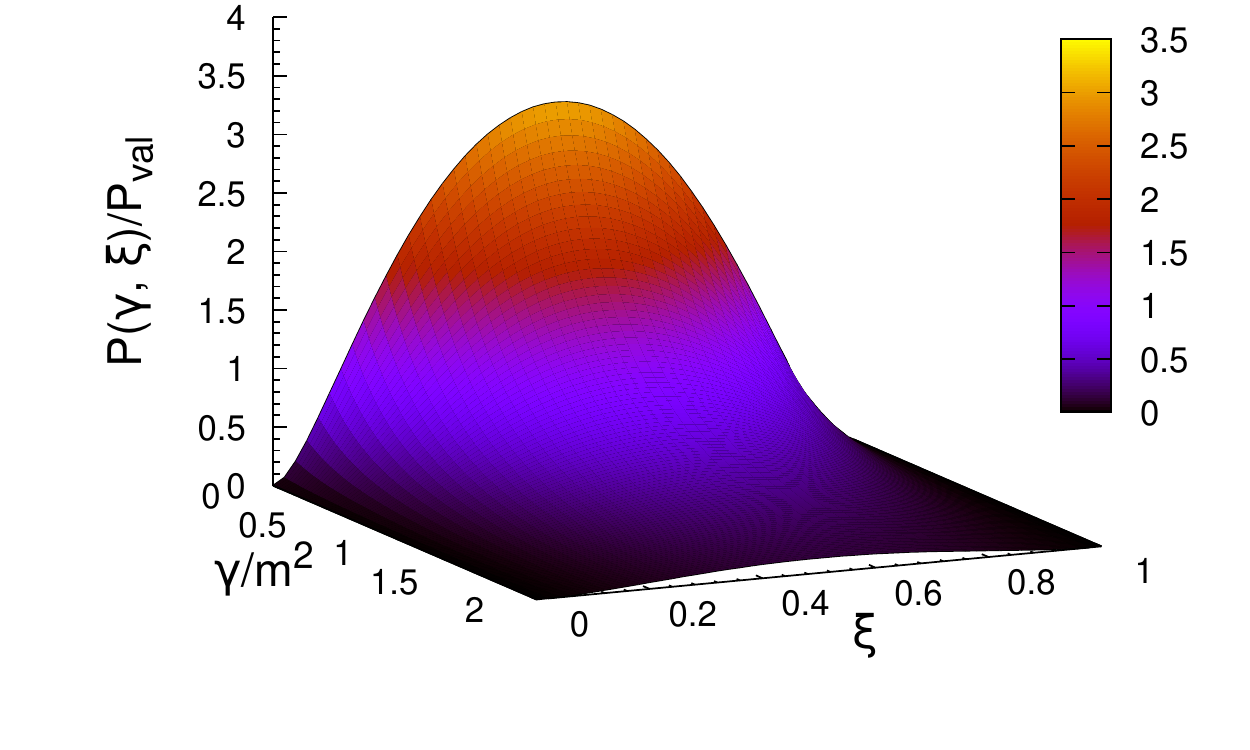,width=8cm}
\epsfig{figure=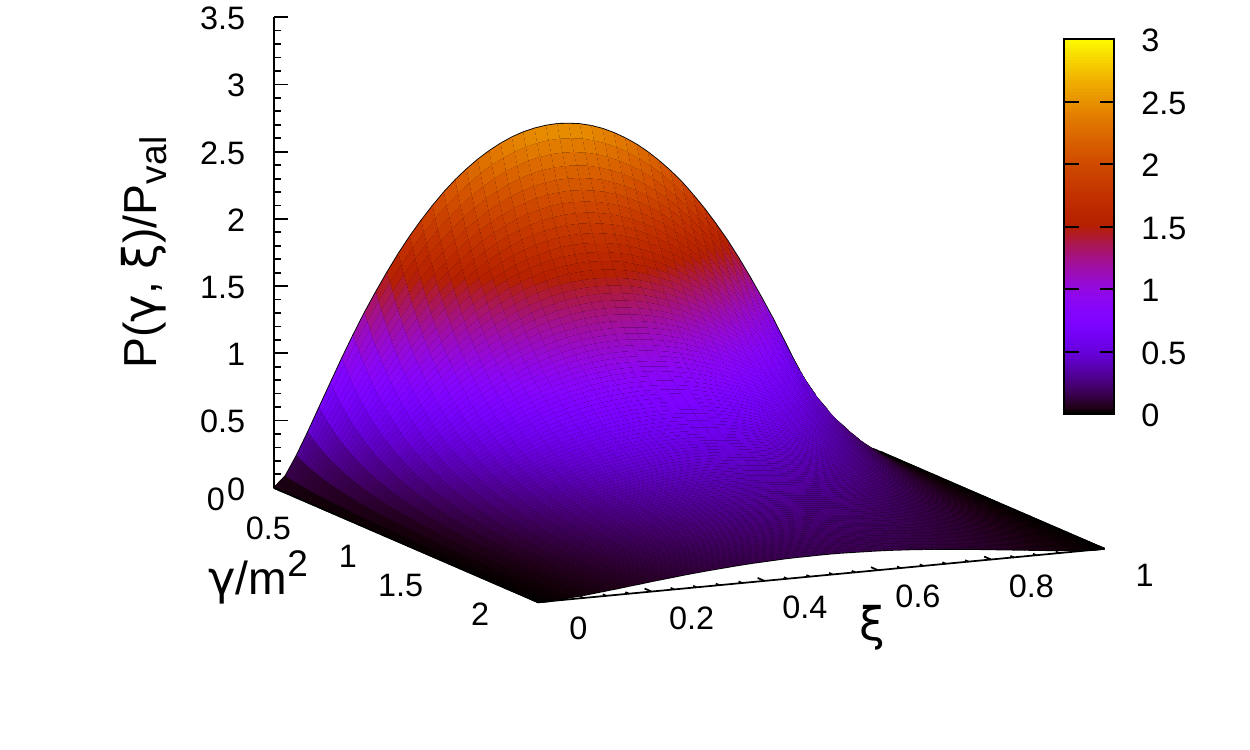,width=8cm}
\caption{\label{fig4} (color online) The LF-momentum density ${\cal P}(\gamma,\xi)$ (see  Eqs. \eqref{pval1},
 \eqref{pval01} and \eqref{pval02}), vs $\xi$ and $\gamma=k^2_\perp$.
From top to bottom:  results for the parameter sets   I,  IV and  VIII, respectively.}
 \end{center}
\end{figure}

Finally, we show the distribution amplitude (DA)~\cite{LepPLB1979,EfrePLB1980,LepPRD1980,Brodsky:1997de}.  This quantity is introduced
through the factorization of the amplitudes associated with exclusive processes and can be expressed as an integral on the transverse-momentum dependence of 
 the valence wave function. In particular, we have evaluated the following spin decompositions 
\be\label{DA}
\varphi_{\uparrow \downarrow} (\xi)=\frac{\int^\infty_0d\gamma ~ \psi_{\uparrow \downarrow} (\gamma,z) }
{\int^1_0 d\xi\int^\infty_0 d\gamma ~ \psi_{\uparrow \downarrow} (\gamma,z)}\, ,\nonu 
\varphi_{\uparrow \uparrow} (\xi)=\frac{\int^\infty_0\,d\gamma ~ \psi_{\uparrow \uparrow} (\gamma,z) }
{\int^1_0 d\xi\int^\infty_0d\gamma ~ \psi_{\uparrow \uparrow} (\gamma,z)}\, .
\ee

\begin{figure}[thb]
\begin{center}
\epsfig{figure=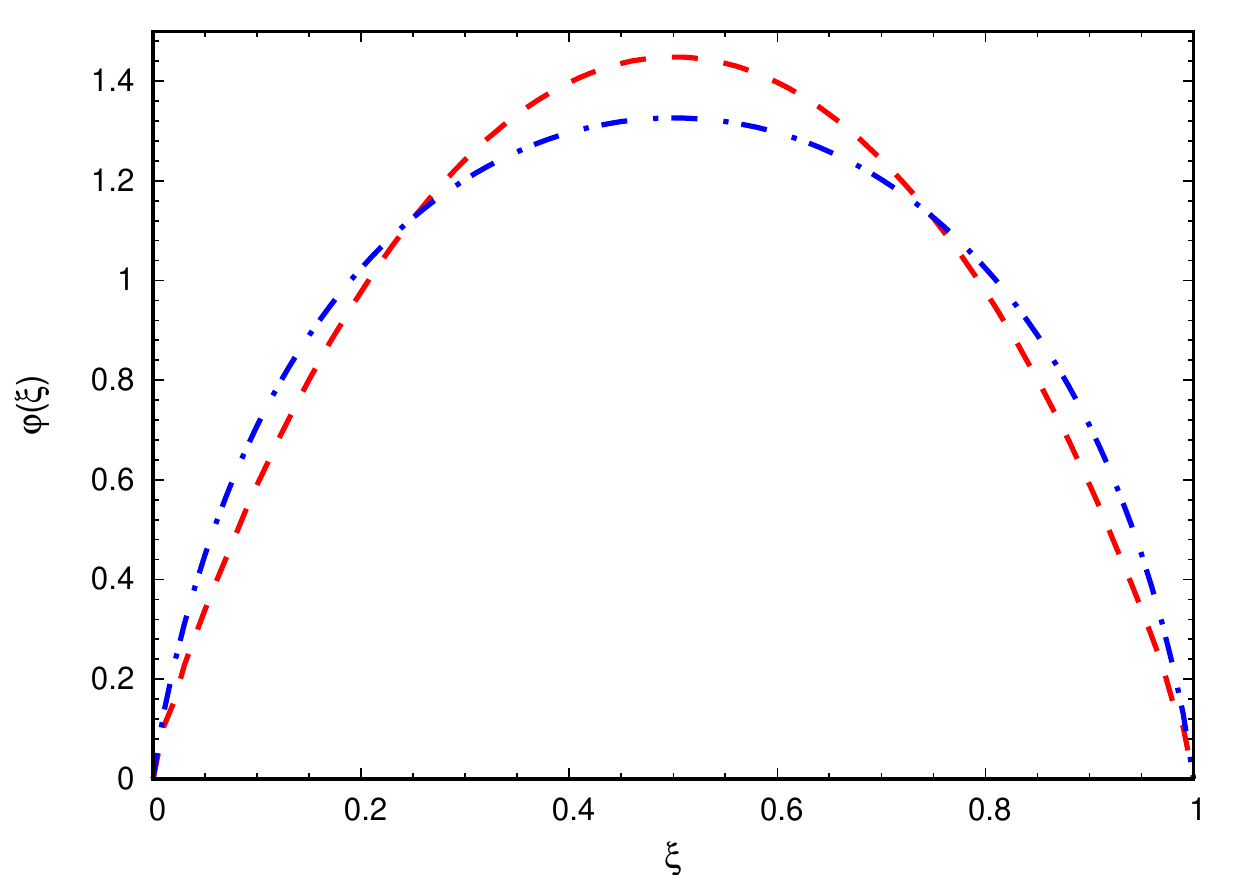,width=8cm}
\epsfig{figure=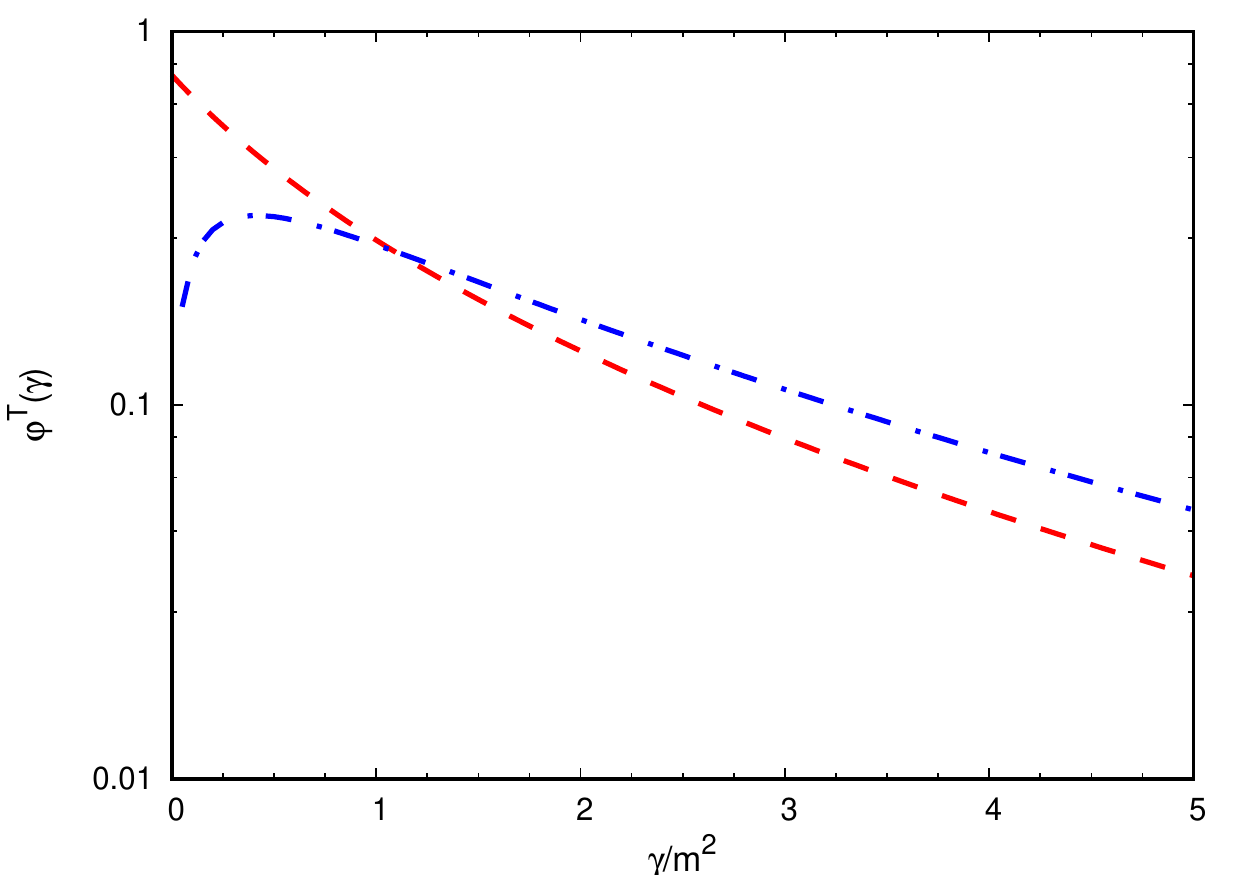,width=8cm}
\caption{(color online) Pion distribution amplitude (upper panel) and the transverse one 
(lower panel) for the two spin components, obtained with  the parameter set VIII. 
Dashed line: anti-aligned, $S=0$, component. Dash-dotted line: aligned, $S=1$, component.}
\label{Fig:DADT_pdf} 
 \end{center}
\end{figure}
 Analogously, we have introduced
the transverse distribution amplitude (TDA) by integrating  the valence  wave
 function over the fraction of longitudinal momentum carried by the valence quark (recall
 $z=1-2\xi$), viz
\be\label{TA}
\varphi^T_{\uparrow \downarrow} (\gamma)=\frac{\int^1_0d\xi \, \psi_{\uparrow \downarrow} (\gamma,z) }{\int^1_0 d\xi\int^\infty_0d\gamma  \, \psi_{\uparrow \downarrow} (\gamma,z)}\,  ,\nonu 
\varphi^T_{\uparrow \uparrow} (\gamma)=\frac{\int^1_0 d\xi\,  \psi_{\uparrow \uparrow} (\gamma,z) }
{\int^1_0 d\xi\int^\infty_0d\gamma \, \psi_{\uparrow \uparrow} (\gamma,z)}\, .
\ee
It has to be pointed out that the TDA is the Fourier transform of  Eq. \eqref{phitildeT}, namely  the transverse amplitude in the 
transverse-coordinates space.
The TDA can be also obtained from Euclidean-space calculations~(see, e.g,. Ref. \cite{Gutierrez:2016ixt}).

The results for the  two spin configurations of both DA and  TDA,   
{ obtained by
using the  parameters of the set VIII},  are shown in 
Fig.~\ref{Fig:DADT_pdf}.
 It is interesting to observe that the aligned component 
of the DA is  wider and decreases slower at the end-points than  the anti-aligned component, as 
it happens for the longitudinal-momentum distributions (c.f.~Table~\ref{table2}). The corresponding features can be 
recognized in the TDA case, where calculations are presented up to  $\gamma/m^2 \sim 5$ (about 0.3 GeV$^2$) showing  the  characteristic IR  scale of  about  $\gamma/m^2 \sim 0.2$,  implicitly carried by our 
 input parameters.  It should be recalled that while  the UV region is governed by 
 the one-gluon exchange, i.e. the
short-range interaction, the IR region
incorporates the features dictated by the long-range correlations in the transverse-coordinate space.

 \subsection{The 3D image of the pion on the null-plane}

\begin{figure}[htb] 
\begin{center}
\epsfig{figure=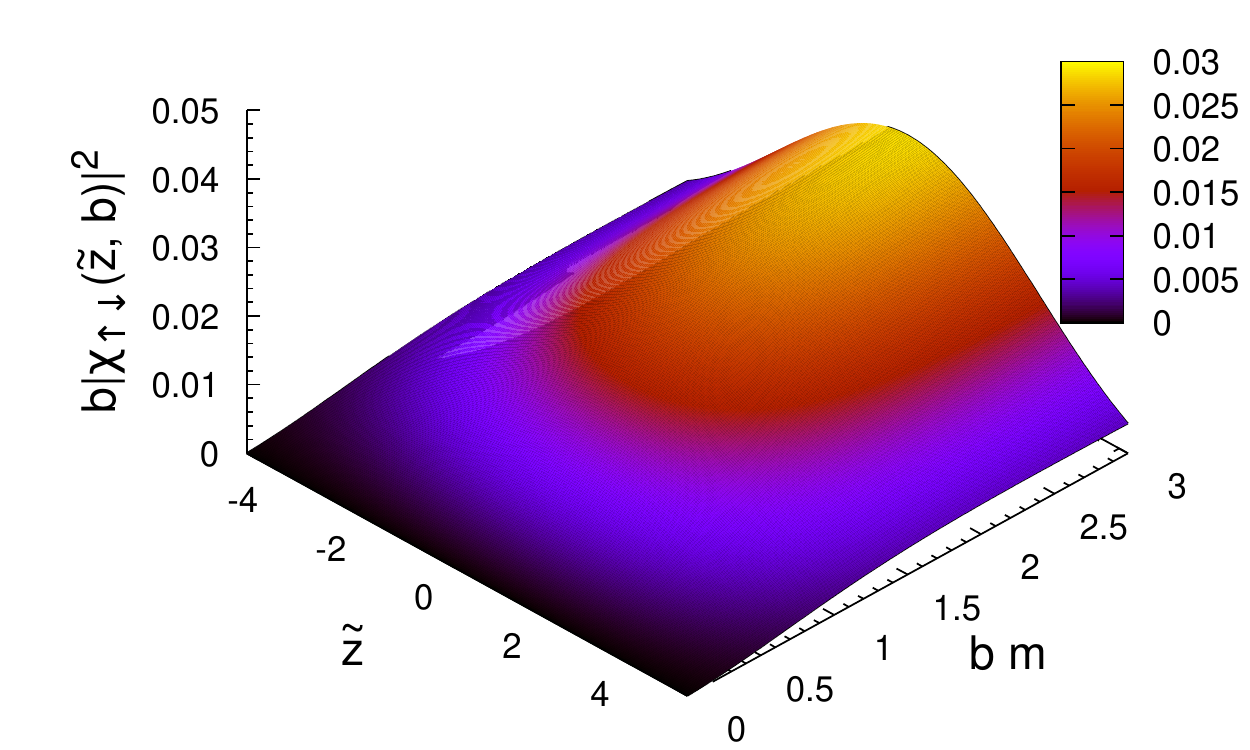,width=8.1cm}
\epsfig{figure=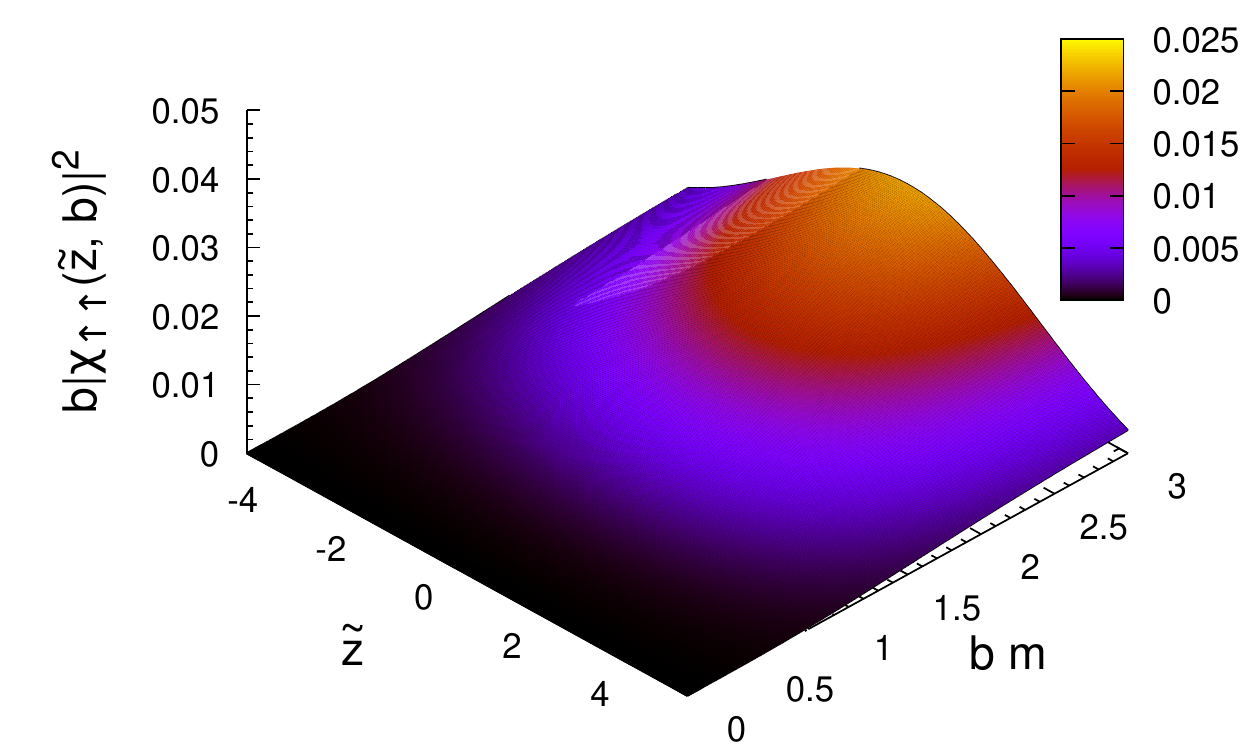, width=8.1cm}
\caption{\label{fig5}  (color online)  The pion in the 3D space $\{\tilde z, {\bf b}\}$ (notice 
the rotational symmetry in the transverse plane) for the set VIII. Upper panel:
anti-aligned component, $b\bigl|\chi_{\uparrow\downarrow}(\tilde z, b)\bigr|^2$.  Lower panel: aligned component,
$b\bigl|\chi_{\uparrow\uparrow}(\tilde z, b)\bigr|^2$.  The variable $b\,m$ has been adopted for making the transverse coordinate adimensional.}
 \end{center}
\end{figure}

 In the 3D space described  by  the Ioffe-time and the transverse coordinates, 
one can obtain an image of the pion  in terms of the
two spin components of the valence wave function, i.e.,
$\chi_{\uparrow\downarrow(\uparrow\uparrow)}(\tilde z, b)$, given in Eq. \eqref{psit2}.
Such a picture of the pion allows one to 
understand better the interplay between short 
and long light-like distances in the description of the hadron structure (see Sec.~\ref{sec_ioffe}). 
In view of this, one   notices  that the region with small values 
of $\{\tilde z,{\bf b}\}$ is  the place where the UV effects should manifest.
Beside the 3D image,  we also present the transverse amplitudes,  $\tilde \varphi^T_{\uparrow\downarrow}(b)$ and
$\tilde \varphi^T_{\uparrow\uparrow}(b)$, given in  Eq.~\eqref{phitildeT}, since they could be the target of 
 LQCD studies.
To perform the  calculations shown in this subsection,  we have used  the parameter set VIII (see
Table~\ref{table1}) that fits  
the pion decay constant.

 The 3D image of the pion spin components on the null-plane is provided in Fig.~\ref{fig5}. Notice that the  exponential factor  
${\rm exp}(-\kappa\, b)$,  present in  the Fourier transform of $\varphi_2(\xi,{\bf k}_\perp,\sigma_i;M,J^\pi,J_z)$ (cf. Eqs.
 \eqref{f0(b,z)},  \eqref{fl0(b,z)} and  \eqref{f1(b,z)})
is  factorized out  in  both   $\chi_{\uparrow\downarrow(\uparrow\uparrow)}(\tilde z, b)$, allowing to use 
a linear scale  in the 3D plot.
 For the purpose of
 the figure, each component is multiplied by the transverse coordinate $b$,  
 canceling
a log-type singularity at $b=0$, generated  by   the Bessel function $K_0$.  

\begin{figure}[htb]
\begin{center}
\epsfig{figure=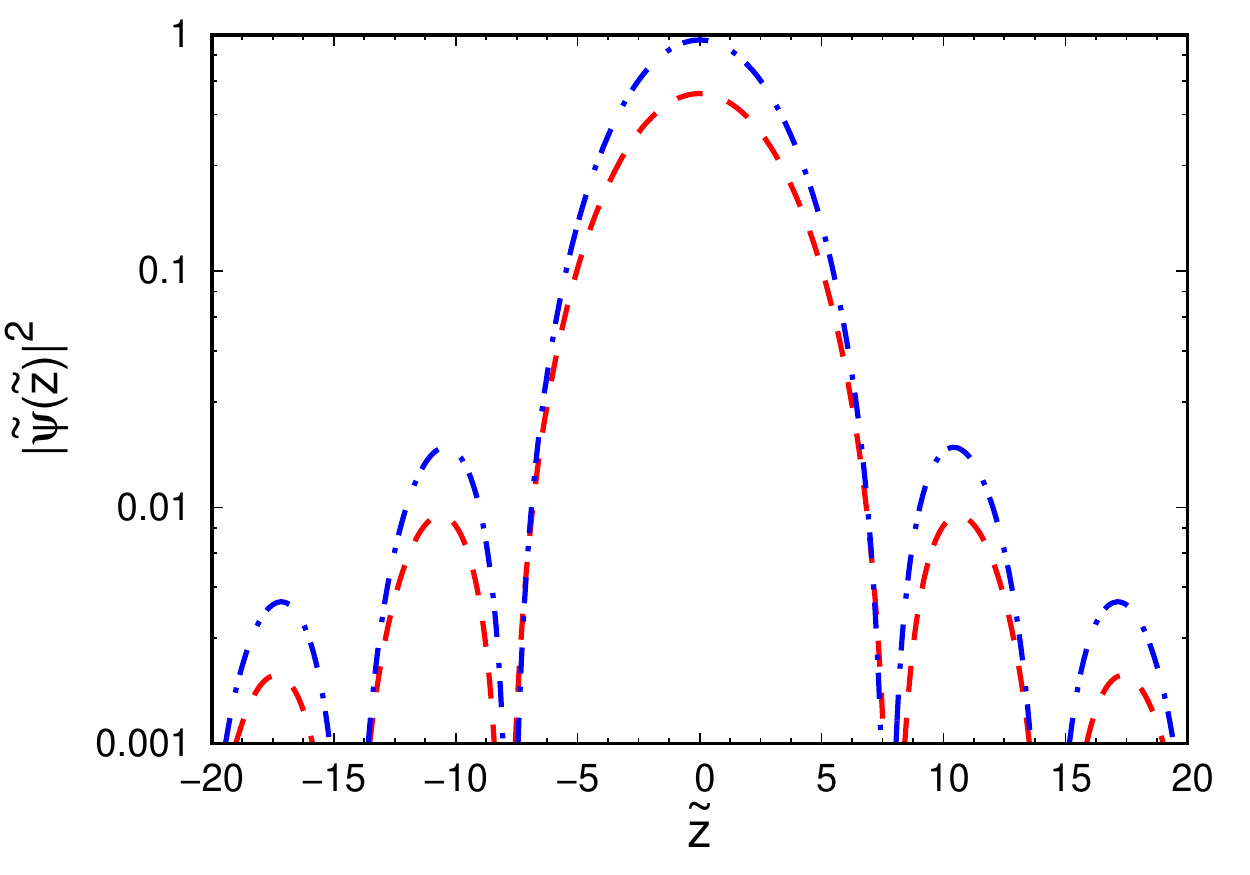,width=8cm}
\caption{\label{fig6} {(color online) Squared integrated amplitudes  $\bigl|
\tilde\Psi_{\uparrow\uparrow(\uparrow\downarrow)}(\tilde z )\bigr|^2$ (cf.  Eq.
\eqref{young}) vs the Ioffe-time, $\tilde z$, by using the parameter set VIII.
 Dashed line: anti-aligned component. Dash-dotted line:
aligned component. }
}
 \end{center}
\end{figure}

A general feature of both densities is the sharp enhancement for 
$\tilde z =0$, i.e., at vanishing light-like distances. Inspired by such an enhancement,
 and in order to better analyze the physically significant dependence  upon $\tilde z$ of the valence 
 wave function,   we have also studied  the  absolute value squared
  of the integrals 
  \be \tilde \Psi_{\uparrow\downarrow(\uparrow\uparrow)}(\tilde z )= {\int_0^\infty{ db \, b}~\tilde\psi_{\uparrow\downarrow(\uparrow\uparrow)}
 (\tilde z, {b} )\over  \int_{0}^\infty {  db \, b}\int_{-\infty}^\infty d\tilde z ~\tilde\psi_{\uparrow\downarrow(\uparrow\uparrow)}
 (\tilde z, {b} )}\, ,
  \label{young}\ee
 with $\tilde\psi_{\uparrow\downarrow(\uparrow\uparrow)}(\tilde z, {b} )$ 
 given in  Eq. \eqref{psitilde}. 
    These amplitudes 
     have a more direct link to the 
 spin  components of
  the valence LF wave function since they also contain the original 
  exponential factor ${\rm e}^{-\kappa\, b}$.  
  
  The integrated amplitudes in Eq.~\eqref{young} {describe} each spin configuration where the constituents are at 
   light-like distance  $\tilde z$ and have relative transverse-momentum 
 ${\bf k}_\perp =0$.  As shown in Fig.~\ref{fig6}, 
  {the amplitudes}
  $\bigl|\tilde\Psi_{\uparrow\downarrow(\uparrow\uparrow)}(\tilde z)\bigr|^2$, have a
   nice   
   diffraction pattern, that  represents a  peculiar feature of 
   such quantities, emphasizing   the interference content due to the entangled
   $q\bar q$ pair.

\begin{figure}[htb]
\begin{center}
\epsfig{figure=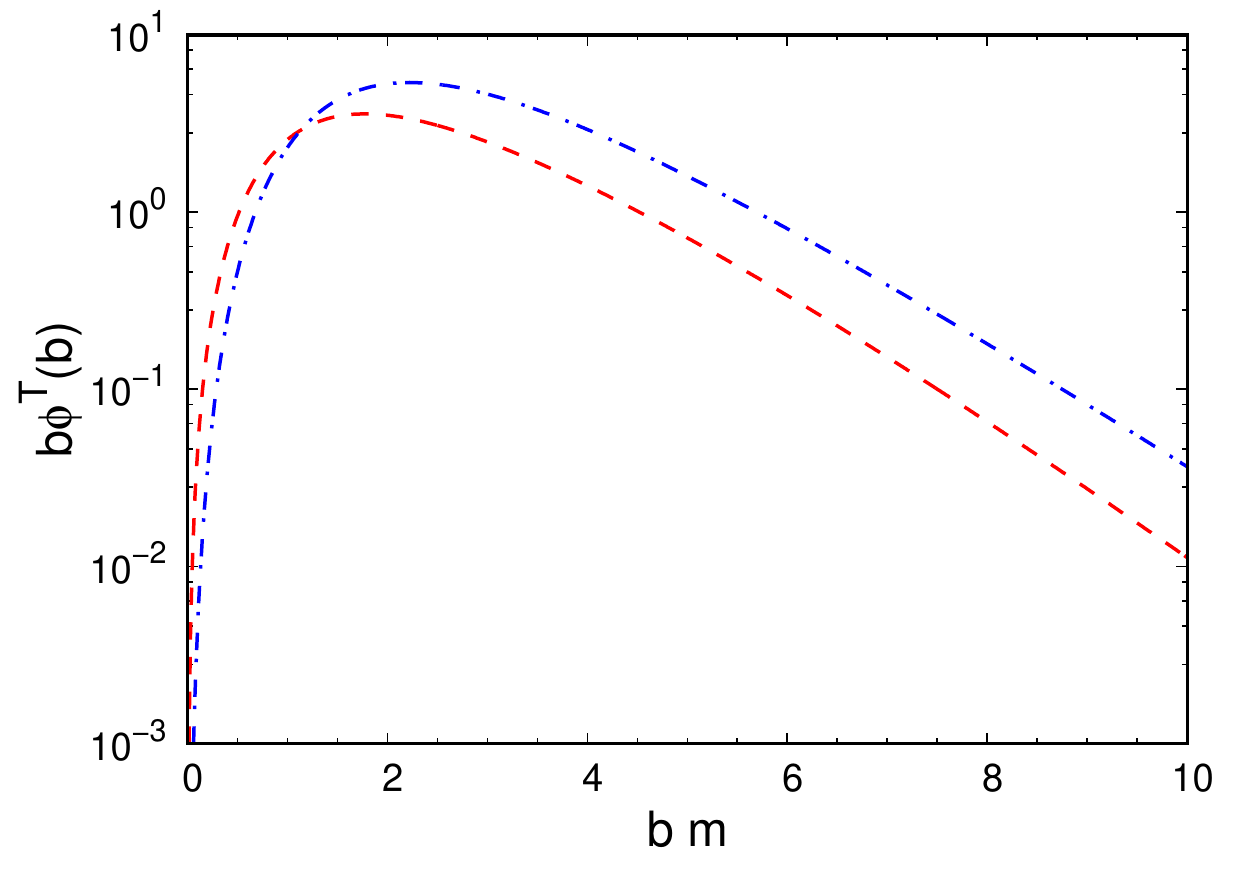,width=8cm}
\caption{\label{fig7} (color color) Transverse amplitudes  for anti-aligned and aligned spin components of 
the pion 
valence wave function as a function  $b\equiv|{\bf b}|$, for the parameter set 
VIII. Dashed line: anti-aligned component. Dash-dotted line: aligned component.} 
 \end{center}
\end{figure}

 Finally, in Fig.~\ref{fig7},  we present the Fourier transform of the 
 transverse amplitudes, $\varphi^T_{\uparrow \downarrow(\uparrow \uparrow)} (\gamma)$ 
 (Eq. \eqref{TA}), 
 in the  transverse-coordinate space, i.e.,
 \be
 \phi^T_{\uparrow \uparrow(\uparrow \downarrow)} (b)={\int_{-\infty}^\infty d\tilde z ~
 \tilde\psi_{\uparrow\downarrow(\uparrow\uparrow)}
 (\tilde z, {b} )\over  \int_{0}^\infty {  db \, b}\int_{-\infty}^\infty d\tilde z ~\tilde\psi_{\uparrow\downarrow(\uparrow\uparrow)}
 (\tilde z, {b} )}\, .
 \ee  
   The log-plot allows one to recognize 
  the characteristic exponential decay at large distances, clearly dominated by   
   $\exp(-\kappa b)$, which   is the familiar fall-off of the wave function
   of a  bound state 
  ($\kappa^2=m^2-M^2/4>0$). Since the amplitudes depend upon transverse
  coordinates, they  should be accessible to LQCD. Therefore, it
  could be interesting to compare its predictions with phenomenological 
  calculations in order to get information about the highly nonlinear
  behavior of QCD at large transverse distances.

 \section{Conclusions and Perspectives}
 
Within the light-front framework, where the physical intuition based on the Fock space expansion
 of the hadron wave function can be used at large extent, 
we have studied  the strongly bound $q\bar q$ system that generates the pion. In our approach,  the 
Bethe-Salpeter equation in Minkowski space is solved by using the Nakanishi integral
representation of the BS amplitude. The ladder approximation of the BS interaction kernel in the Feynman gauge has been considered, 
with the constituent quarks 
interacting by an effective massive gluon exchange.  An extended effective 
quark-gluon vertex function 
is introduced through a form factor, { that contains a new scale parameter,
 beside 
the gluon mass. The  range   where (i) the form-factor parameter 
$\Lambda$, (ii) the quark effective mass and  (iii) the gluon one}  can vary has been chosen according to
 LQCD results, with  the guideline given by the value of the IR scale $\Lambda_{QCD}$.
We fine-tuned the parameters around such a scale in order to reproduce $f_\pi$.
Once the BSE has been solved, we  formally obtained the valence LF wave  function,
that    is well-defined in terms of the lowest number of fields associated to the BS
amplitude, and therefore  it contains unambiguous information on the dynamics  that  one
can convey into the interaction kernel of the BSE itself.

 The detailed study of
the  valence Fock component
 has been carried out  by   first addressing 
 static (or  integral) quantities, like: (i) the pion decay constant, $f_\pi$,  and (ii) the
 valence probability,  whose value is around $70\%$, according to    our calculations. A relevant and peculiar feature of our approach is  the
possibility of decomposing the valence component in the two allowed spin configuration: the dominant $S=0$ component and the purely
relativistic $S=1$, that  yields  a $\sim 20\% $ contribution to the valence probability. Such a decomposition also has been applied to 
the second set of investigated quantities,  bringing a considerable wealth of dynamical information. In particular, we have analyzed:   
(i)  the longitudinal and transverse LF-momentum
distributions; 
(ii)  the distribution amplitudes that depend upon the longitudinal and   transverse LF-momentum components; 
(iii) the probability
densities in momentum and  configuration spaces (the last one is given by the  Cartesian  product of the covariant Ioffe-time and
the Euclidean transverse coordinates).
For each quantity, we have highlighted
  signatures of the  dynamics
governed by the one-gluon exchange and  have also emphasized the relevance of the transverse degrees of freedom, more accessible to the
LQCD studies. 

Future developments of our approach are primarily related to the implementation of both quark and gluon dressed propagators, 
more realistic dressed quark-gluon vertex,
in order to extend to
the Minkowski space the
successful studies  
of the spontaneously broken chiral symmetry performed in the Euclidean space. In this way, the kernel of the BSE 
can be improved systematically, adding step by step new dynamical contents from QCD, that one should explore
 in 
the elaboration of hadron models. 
Furthermore, observables like the electromagnetic form factor and 
{ generalized transverse-momentum  dependent parton distributions
(see, e.g., Ref.\cite{Lorce:2011dv}) } of the pion are within our plans for future studies.

 \begin{acknowledgments}
 
 WdP and TF acknowledge the
warm hospitality of INFN Sezione di Roma.  WdP acknowledges the partial support of CNPQ under grants 438562/2018-6 and 313236/2018-6, 
and the partial support of CAPES under grant 88881.309870/2018-01. TF thank the financial support from 
the Brazilian Institutions: CNPq  (Grant No. 308486/2015-3), CAPES (Finance Code 001),  and FAPESP
(Grants no.~13/26258-4 and 17/05660-0). JHAN acknowledges the support of FAPESP Grant no. 2014/19094-8. 
GS thanks the partial support
of CAPES and acknowledges the warm hospitality of the Instituto 
Tecnol\'ogico de Aeron\'autica. EY acknowledges the support of FAPESP Grants no.~2016/25143 and 2018/21758-2.
This work is a part of the project INCT-FNA Proc.~No.~464898/2014-5.
\end{acknowledgments} 

\newpage
 
  \appendix
 \section{The valence wave function and the valence probability}
 \label{app_val}
 
 In this Appendix, the valence probability is decomposed into the 
 contributions generated by the spin configurations
 active inside the pion. 
 
 Let us first illustrate  the relation between the BS amplitude and the valence 
 wave function, i.e. the amplitude with the
 lowest number of constituents in the  Fock   expansion 
 of the bound system.
 Recalling the key observation contained in Ref.
\cite{YanII}, i.e. the  independent degrees of freedom described by the
fermionic field on the null-plane is  given by 
the projections 
 $ \psi^{(+)}(\tilde x,x^+=0^+)=\Lambda^+ \psi (\tilde x,x^+=0^+)$ 
 with $\tilde
x\equiv\{x^-, {\bf x}_\perp\}$ and 
\be\Lambda^\pm={1 \over 4} \gamma^\mp\gamma^\pm~~,
\quad
\Lambda ^\pm \Lambda ^\pm =\Lambda ^\pm~~, \quad \Lambda^++\Lambda^-=1
~~,
\nonu~ \Lambda^{\pm} \gamma^0\Lambda^{\pm}=0~~, \quad  \Lambda^{\pm}
\gamma^0\Lambda^{\mp}={\gamma^\mp\over 2}~~, \quad
Tr\Bigl\{\Lambda^\pm\Bigr\}=2~~.\nonu
\ee 
one can write  the {\em good component} of the fermion field as
follows
\be
 \psi^{(+)}(\tilde x,x^+=0^+)= 
 \int {d\tilde q\over (2\pi)^{3/2}} ~{\theta(q^+) \over \sqrt{2q^+}}~
 \sum_\sigma \nonu
\Bigl [ U^{(+)}(\tilde q,\sigma)~b(\tilde q,\sigma) e^{i\tilde q \cdot \tilde x} +
  V^{(+)}(\tilde q,\sigma)~d^\dagger(\tilde q,\sigma)e^{-i\tilde q \cdot \tilde x} \Bigr ]
~,\nonu \ee
 where  
 \be U^{(+)}(\tilde q,\sigma)=\Lambda^+ u(\tilde q,\sigma)
 ~~, \quad  V^{(+)}(\tilde q,\sigma)=\Lambda^+ v(\tilde q,\sigma)~,
 \nonu\ee
 with the normalization   $\bar u~u=2m$ 
 and $\bar v(\tilde q,\sigma') v(\tilde q,\sigma)=-2m$. 
 Hence, one  expresses the fermion creation and annihilation operators associated with $b$ and $d$, in terms of the {\em good} component
of the field, as follows
 \be 
 \label{bd}
 (2\pi)^{3/2} ~\theta(q^+) ~\sqrt{2q^+}~b(\tilde q,\sigma')=
 \nonu
 =
\int d\tilde x ~e^{-i \tilde q \cdot \tilde x}
  ~ U^{(+)\dagger}(\tilde q,\sigma') 
  \psi^{(+)}(\tilde x,x^+=0^+)~~,
  \nonu
(2\pi)^{3/2} ~\theta(q^+)~\sqrt{2 q^+}~d^\dagger(\tilde q,\sigma') =
\nonu
=\int d\tilde x ~e^{i \tilde q \cdot \tilde x} ~
  V^{(+)\dagger}(\tilde q,\sigma') 
  \psi^{(+)}(\tilde x,x^+=0^+)  ~~, 
    \ee
where 
\be 
 u^\dagger(\tilde q,\sigma') \Lambda^+ u(\tilde q,\sigma)= q^+
\delta_{\sigma',\sigma}~~, \nonu
 v^\dagger(\tilde q,\sigma') \Lambda^+ v(\tilde q,\sigma)=
q^+ \delta_{\sigma,
 \sigma'}~~.\ee
  Once the creation and
 annihilation operators for the fermions are defined, one can construct
 the Fock component  with the lowest
 number of constituents and therefore introduce the LF valence amplitude, 
 $\varphi_{2}(\xi, \bm{k}_{\perp
 },\sigma_i;M,J^\pi,J_z)$. It reads
{
 \be
 \varphi_{2}(\xi, \bm{k}_{\perp },\sigma_i;M,J^\pi,J_z) =
  (2\pi)^3 ~\sqrt{N_c}~2p^+ \sqrt{\xi(1-\xi)}
\nonu
 \times ~\langle 0| b(\tilde q_2, \sigma_2)
~d(\tilde q_1, \sigma_1)| 
\tilde p, M,J^\pi,J_z\rangle~,
\label{app_valdef}
 \ee
} 
{ where $\tilde q_1\equiv \{q^+_1=M (1-\xi),-{\bf k}_\perp\}$,  
$\tilde q_2\equiv \{q_2^+=M\xi,{\bf k}_\perp\}$ and $\xi= 1/2 +k^+/p^+$}.

 From the general expression of the BS amplitude (cf. Eq.~\eqref{bsa0}, adapted 
 to the general notation of this Appendix) and Eq.~\eqref{app_valdef}, 
 one can show that the valence amplitude is related to the BS one by
\be
 \varphi_{2}(\xi, \bm{k}_{\perp },\sigma_i;M,J^\pi,J_z) 
=
 {\sqrt{N_c} \over p^+} {1 \over 4}~\bar u_\alpha(\tilde q_2,\sigma_2)
 \nonu \times~
\int {dk^{-}\over 2\pi} \Bigl[\gamma^+~\Phi(k,p) \,  \gamma^+
\Bigr]_{\alpha\beta}~
  v_\beta(\tilde q_1,\sigma_1) ~~.
\label{app_valbse}
 \ee 
More explicitly, from Eqs. \eqref{bsa} and \eqref{S_structure} one gets 
  \be
\varphi_{2}(\xi, \bm{k}_{\perp },\sigma_i;M,J^\pi,J_z) =
 {\sqrt{N_c} \over p^+} {1 \over 4}~\bar u(\tilde q_2,\sigma_2)~
\int {dk^{-}\over 2\pi}
\nonu \times~\gamma^+~
~\Biggl[  {\psla p\over M}  ~ \phi_2 +
 \Bigl({k \cdot p \over M^3}  \psla p  - {\psla k\over M} \Bigr) ~ \phi_3 - 
 {1 \over M^2} \psla  p~  \psla k ~  \phi_4 \Biggr] ~\gamma_5
  \nonu\times ~  \gamma^+ ~
  v(\tilde q_1,\sigma_1)=
  \nonu
  =
  -~{\sqrt{N_c} \over p^+} ~\int {dk^{-}\over 2\pi}
 ~
\Biggl\{  \Biggl[ \phi_2 +
 \Bigl({k^- \over 2M}+{z\over 4}\Bigr)    ~ \phi_3\Biggr]
 \nonu \times~ D_1^{\sigma_1\sigma_2}(\tilde q_1,\tilde q_2)
 -
 {1 \over M} ~ 
 D_2^{\sigma_1\sigma_2}(\tilde q_1,\tilde q_2)
  ~  \phi_4 \Biggr\} 
 ~~,
\label{app_spincont}\ee
where the following relations have been exploited: $\gamma^\pm  \gamma^\pm=0$, 
 $\gamma^-~\gamma^+~\gamma^-=4\gamma^-$ (and the analogous for the other combination). Moreover,
 \be
 D_1^{\sigma_1\sigma_2}(\tilde q_1,\tilde q_2)=
 ~Tr\Bigl\{{\cal P}^{\sigma_1\sigma_2}(\tilde q_1,\tilde q_2)~\gamma_5~
 \Lambda^+\Bigr\}  
  \nonu
  D_2^{\sigma_1\sigma_2}(\tilde q_1,\tilde q_2)=
 ~Tr\Bigl\{{\cal P}^{\sigma_1\sigma_2}(\tilde q_1,\tilde q_2)
  ~{\bf k}_\perp \cdot {\bm \gamma}_\perp
  ~\gamma_5~
  \Lambda^+\Bigr\}~,
  \nonu
  \label{D1_D2}
 \ee
 with 
 \be
 {\cal P}^{\sigma_1\sigma_2}(\tilde q_1,\tilde q_2)=v(\tilde q_1,\sigma_1)u^\dagger(\tilde q_2,\sigma_2)
~~.\ee
The  spinors $u$ and $v$ in terms of the LF variables can be found 
 in Appendix  B of Ref.~\cite{Brodsky:1997de},  namely
\be
u(\tilde q,\sigma)
={ q^+ +\gamma^0~m +
\gamma^0{\bf q}_\perp\cdot{\bm \gamma}_\perp\over  \sqrt{ 2 q^+}}~\begin{pmatrix} \chi^\sigma \\
\sigma \chi^\sigma \end{pmatrix}
\ee
and
\be
v(\tilde q,-\sigma)=
  { q^+ -\gamma^0~m +\gamma^0{\bf
q}_\perp\cdot{\bm \gamma}_\perp\over  \sqrt{ 2 q^+}} ~\left(\begin{array} {c} 
\chi^{\sigma} \\
 \sigma\chi^{\sigma} \end{array}\right)~~,
 \nonu
 \ee
 with $\sigma=\pm 1$. Notice that in $v$ there is an opposite sign for the helicity (in
the LF framework helicity and third component of the spin coincide, see, e.g.,
\cite{Chiu}). For  a quick evaluation of the above matrix elements in Eqs.~\eqref{D1_D2}, it is 
 useful to introduce  the following dyadic products
 \be
\begin{pmatrix}\chi^\sigma \\ \sigma\chi^\sigma \end{pmatrix} \begin{array}{c}
\begin{pmatrix}\chi^{ \sigma\dagger} & \sigma\chi^{ \sigma\dagger} \end{pmatrix}\\
~\end{array}=
(1+\sigma\gamma_5)~\Lambda^+~~,
\label{app_dyad1}
\ee \be 
\begin{pmatrix}\chi^{-\sigma} \\ -\sigma\chi^{-\sigma} \end{pmatrix} \begin{array}{c}
\begin{pmatrix}\chi^{ \sigma\dagger} & \sigma\chi^{ \sigma\dagger} \end{pmatrix}\\
~\end{array}=
(1-\sigma\gamma_5)~{\gamma_{L(R)}\over \sqrt{2}}~~,
\nonu \label{app_dyad2}\ee
where {in Eq. \eqref{app_dyad2} for  $\sigma= \pm 1$ one has to use 
$\gamma_{L(R)}$, with}
 \be
 \gamma_{R(L)}= \mp ~{1\over \sqrt{2}}(\gamma^1\pm i\gamma^2)
 ~~.  
 \ee
The actual expressions  of $D_1$ and $D_2$ can be obtained through
 suitable  traces. In particular, for $D_1^{-\sigma\sigma}$ one gets
\be
D_1^{-\sigma\sigma}(\tilde q_1,\tilde q_2)=
 Tr\Bigl\{ { q^+_1 -\gamma^0~m  { -} \gamma^0{\bf
k}_\perp\cdot{\bm \gamma}_\perp\over  \sqrt{ 2 q^+_1}}
\nonu \times ~(1+\sigma\gamma_5)
\Lambda^+~{ q^+_2 +\gamma^0~m { +}
\gamma^0{\bf k}_\perp\cdot{\bm \gamma}_\perp\over  \sqrt{ 2 q^+_2}}
 ~\gamma_5~\Lambda^+\Bigr\}=
 \nonu
={ \sqrt{q^+_1q^+_2}\over 2}
 ~Tr\Bigl\{ (1+\sigma\gamma_5)
 ~\Lambda^+~
 ~\gamma_5~\Lambda^+\Bigr\} =
 \nonu
 = \sigma~{M}\sqrt{\xi(1-\xi)} = \sigma~ {M\over 2}\sqrt{1-z^2}~~,
\ee
where  we can properly move the projector $\Lambda^+$ inside the trace 
 for simplifying the whole expression, e.g.  
\be
\Lambda^+~[q^+_1 -\gamma^0~m -\gamma^0{\bf
k}_\perp\cdot{\bm \gamma}_\perp]\Lambda^+=\Lambda^+~q^+_1 
\nonu
\Lambda^+~[q^+_2 +\gamma^0~m +\gamma^0
{\bf k}_\perp\cdot{\bm \gamma}_\perp]\Lambda^+=\Lambda^+~q^+_2 ~.
\ee
 By using the
previous relations, one can show that the matrix  $D_2^{-\sigma\sigma}$
 vanishes, viz.
\be
D_2^{-\sigma\sigma}(\tilde q_1,\tilde q_2)=
 Tr\Bigl\{ { q^+_1 -\gamma^0~m  { -} \gamma^0{\bf
k}_\perp\cdot{\bm \gamma}_\perp\over  \sqrt{ 2 q^+_1}}
\nonu \times ~(1+\sigma\gamma_5)
~\Lambda^+
~{ q^+_2 +\gamma^0~m { +}
\gamma^0{\bf k}_\perp\cdot{\bm \gamma}_\perp\over  \sqrt{ 2 q^+_2}}
 \nonu \times~{\bf
q}_\perp\cdot{\bm \gamma}_\perp~
  \gamma_5~
  \Lambda^+\Bigr\}=0 ~~.
\ee
Also $D_1^{\sigma\sigma}$ is vanishing, once the following relation is adopted 
\be
\gamma_5~[q^+_2 -\gamma^0~m +
\gamma^0{\bf k}_\perp\cdot{\bm \gamma}_\perp]\Lambda^+
\nonu \times~[q^+_1 -\gamma^0~m -\gamma^0{\bf
k}_\perp\cdot{\bm \gamma}_\perp]=
\nonu
=\gamma_5~q^+_2[q^+_1\Lambda^+ -{\gamma^-\over 2}~m -{\gamma^-\over 2}{\bf
k}_\perp\cdot{\bm \gamma}_\perp] 
\nonu-\gamma_5~{m\over 2} [q^+_1\gamma^+- \gamma^+{\gamma^-\over 2}~m
]
\nonu
 +\gamma_5~{1\over 2}{\bf
k}_\perp\cdot{\bm \gamma}_\perp
[q^+_1\gamma^+  -\gamma^+{\gamma^-\over 2}{\bf
k}_\perp\cdot{\bm \gamma}_\perp]~,
\ee
and the standard results for the traces of the Dirac matrixes. Namely, one gets
\be
D_1^{\sigma\sigma}(\tilde q_1,\tilde q_2)=
 Tr\Bigl\{ { q^+_1 -\gamma^0~m { -} \gamma^0{\bf
k}_\perp\cdot{\bm \gamma}_\perp\over  \sqrt{ 2 q^+_1}}
\nonu \times ~(1-\sigma\gamma_5)
{\gamma_{L(R)}\over \sqrt{2}}~~{ q^+_2 +\gamma^0~m { +}
\gamma^0{\bf k}_\perp\cdot{\bm \gamma}_\perp\over  \sqrt{ 2 q^+_2}}
 \nonu\times~\gamma_5~\Lambda^+\Bigr\}=0~~.
\ee
The last matrix element $D_2^{\sigma\sigma}$ is given by 
\be
D_2^{\sigma\sigma}(\tilde q_1,\tilde q_2)=
 Tr\Bigl\{ { q^+_1 - \gamma^0~m { -}\gamma^0
 {\bf k}_\perp\cdot{\bm \gamma}_\perp\over  \sqrt{ 2 q^+_1}}
\nonu \times ~(1-\sigma\gamma_5)
~{\gamma_{L(R)} \over \sqrt{2}}
~{ (q^+_2 +\gamma^0~m ){\bf k}_\perp\cdot{\bm \gamma}_\perp { +}
\gamma^0|{\bf k}_\perp|^2\over  \sqrt{ 2 q^+_2}}
 \nonu \times~
  \gamma_5~
  \Lambda^+\Bigr\}  
  =-{\sqrt{q^+_1 q^+_2}\over 2\sqrt{2}}\nonu\times ~Tr\Bigl\{  (1-\sigma\gamma_5)
   ~\gamma_{L(R)}~\Bigl(k_R \gamma_L +k_L \gamma_R \Bigr)~\gamma_5\Lambda^+
   \Bigr\}=
   \nonu
   =\sigma
   ~k_{L(R)}M~\sqrt{(1-z^2)\over 2}~~,
  \ee
  where 
  \be\label{eq:klr}
  k_{R(L)}= \mp ~{1\over \sqrt{2}}(k_x  \pm i k_y)~~,
  \ee 
  and the following relations have been used
  \be\gamma_{R(L)}\gamma_{R(L)}=0~~,
  \nonu
  {1\over 4} Tr\Bigl\{ (\gamma_5-\sigma) 
~\gamma^-\gamma^+ \gamma_{L(R)}\gamma_{R(L)} 
 \Bigr\} = 
 \nonu
 ={1\over 2} Tr\Bigl\{ (\gamma_5-\sigma) 
~(1+\gamma^0\gamma^3)  ~(1 -\sigma~\sigma^{12})
 \Bigr\}=
 \nonu
 = -2 \sigma - 2\sigma~{1\over 4}  Tr\Bigl\{ \gamma_5 
~\gamma^0\gamma^3  ~i\gamma^1 \gamma^2
 \Bigr\}=
 \nonu
 =-4 \sigma~~,
  \ee
  { recalling that for $\sigma= \pm 1$ one has the product 
  $\gamma_{L(R)}~\gamma_{R(L)}$.}
 Finally, by  inserting the above results into Eq. \eqref{app_spincont}, one gets 
  \be\varphi_{2}(\xi, \bm{k}_{\perp },\sigma_i;M,J^\pi,J_z) 
  =
~- {\sigma_2\over 2} ~\sqrt{N_c}~\sqrt{1-z^2}
\nonu
\times~\int {dk^{-}\over 2\pi}
\Biggl\{  \delta_{\sigma_2,-\sigma_1}\Biggl( \phi_2(k,p) +
 \Bigl({k^- \over 2M}+{z\over 4}\Bigr)     \phi_3(k,p)\Biggr)  
  \nonu {{-}}
 { k_{L(R)}\sqrt{2} \over  M} ~\delta_{\sigma_2,\sigma_1} ~  \phi_4(k,p)\Biggr\}
~~.
\label{app_spincont2}\ee
The integration over $k^-$ can be readily performed by using the NIR of the
scalar functions $\phi_i$, Eq. \eqref{phinak}. Explicitly, one gets (cf
 Refs. \cite{dePaula:2016oct,dePaula:2017ikc})
\be
{{\psi_i(\gamma,\xi)}}= -{i\over M}~\int_0^\infty d\gamma'
\nonu \times~{g_i(\gamma',z;\kappa^2)\over 
[\gamma'+\gamma+m^2z^2 +(1-z^2) \kappa^2-i\epsilon]^2}~~,
\label{app_psi_i}
\ee
with $z=1-2\xi$.
Hence, one can write
\be
\varphi_{2}(\xi, \bm{k}_{\perp },\sigma_i;M,J^\pi,J_z) 
  =
~- {\sigma_2\over 2} ~\sqrt{N_c}~\sqrt{1-z^2}
\nonu
\times
~\Biggl\{  \delta_{\sigma_2,-\sigma_1}
\Biggl[ \psi_2(\gamma,z) +{z\over 2} {{\psi_3(\gamma,z)}}
\nonu+ {i\over M^3} 
~\int_0^{\infty} d\gamma'~
~{\partial g_{3}(\gamma',z;\kappa^2)/\partial z\over
  \gamma+\gamma'+z^2m^2 +(1-z^2)\kappa^2 -i \epsilon} ~ 
  \Biggr]
  \nonu ~-~
 { k_{L(R)}\sqrt{2} \over  M} ~\delta_{\sigma_2,\sigma_1} ~  \psi_4(\gamma,z)\Biggr\}
~~ ,
\label{app_spincont3}
\ee 
where { for $\sigma_2=\sigma_1=\pm 1$ one has to use $k_{L(R)}$.}
From  Refs. \cite{dePaula:2016oct,dePaula:2017ikc}, the cumbersome term $k^- ~\phi_3$ contains a LF singularity, 
  can be
manipulated as follows
\be
\int {dk^{-}\over 2\pi}
~ k^- ~ \phi_3(k,p)=
\nonu
=\int_{-1}^{+1} dz' \int_0^{\infty} d\gamma' g_{3}(\gamma',z';\kappa^2)
~{\cal I}_1(\gamma',z',z)~~,
\label{app_LF_sing}
\ee
where
\be
{\cal I}_1(\gamma',z',z)=\int {dk^- \over 2\pi}\nonu\times
~{k^- \over 
[k^-(z'-z){M\over 2}  -zz'{M^2\over 4} -\gamma-\gamma' -\kappa^2 +i \epsilon]^3}
~,\nonu\ee
with $z=-2 k^+/M$. N.B. the integral ${\cal I}_1$ becomes singular for $z'=z$. Its
evaluation proceeds by using (see also Ref. \cite{YanIV})
\be
\int_{-\infty}^\infty {dx\over 2\pi} ~
{ 1 \over \Bigl[\beta~x -y\mp i \epsilon\Bigr]^n}= ~\pm {i\over n-1}
~{\delta (\beta)\over \Bigl[-y\mp i \epsilon\Bigr]^{n-1}}~~.
\nonu\ee
Thus, one gets
\be
{\cal I}_1(\gamma',z',z)=  -~ i {2\over M^2}
{\delta(z'-z)\over (z'-z)} 
\nonu\times~{1\over
\left [-zz'M^2/4 -\gamma-\gamma' -\kappa^2 +i \epsilon\right]}=\nonu
= i~\delta'(z'-z)~
{2\over M^2} ~{1\over
\left [-zz'M^2/4 -\gamma-\gamma' -\kappa^2 +i \epsilon\right]}~,
\nonu\ee
where the derivative of the delta function fulfills the relation:
 $x~\delta'(x)=-\delta(x)$,
since $d[x~\delta(x)]/dx=0$. By inserting this result into Eq. 
\eqref{app_LF_sing}, and properly integrating by parts, one gets
\be
\int {dk^{-}\over 2\pi}
~ {k^-\over 2M} ~ \phi_3(k,p)=
\nonu
={i\over M^3} 
~\int_0^{\infty} d\gamma'~
~{\partial g_{3}(\gamma',z;\kappa^2)/ \partial z\over
\left [\gamma+\gamma'+z^2m^2 +(1-z^2)\kappa^2 -i \epsilon\right]} 
\nonu+  {z \over 4}~\psi_{3}(\gamma,\xi) {\color{blue}.}
\ee
\section{Orbital angular momentum decomposition on the transverse-plane}
\label{app_OAM}
 In this Appendix we illustrate some formal steps for obtaining the Fourier transform of the 
  the two 
 spin components of  the
 valence wave function $\varphi_2$ (cf. Eq. \eqref{spin_cont2}), as given in Eq. \eqref{phibgm}.
   As a by-product, the obtained  results 
 allow one to 
 emphasize the $L_z$ content in  each contribution (recall that 
  $L_z=xp_y-yp_x$, and  it is a kinematical
  generator, within the LF framework).    
  
  Let us first recall that 
\begin{equation}
{\rm e}^{{i} x \cos\theta}=\sum_{m=-\infty}^{m=\infty}{i}^m J _m(x) {\rm e}^{{i} m\,\theta} \, ,
\end{equation}
where $J_m(x)$ is the Bessel function of integer order.
 In particular,  the  Fourier transform of the spin anti-aligned and  aligned components 
  are associated with integrals of $J_0$ and $J_1$ respectively, after performing the angular
   integration in Eq. \eqref{phibgm}.

By adopting the expression of $\psi_{\uparrow\downarrow}(\gamma,z)$ and  
$\psi_{\uparrow\downarrow}(\gamma,z)$ in terms of the respective NIRs (cf. 
 Eqs.  \eqref{psiantiparallel} 
and \eqref{psiparallel}), one has to evaluate the following integrals.
The first one is 
\begin{multline}\label{f0(b,z)}
F_0(z,\gamma',b)=\int^\infty_0 d\gamma~{J_0(b \sqrt{\gamma})
\over \gamma+\gamma' +\kappa^2+z^2\tfrac{M^2}{4}} \\ =2K_0\left(b\,\sqrt{\gamma' +\kappa^2+z^2\tfrac{M^2}{4}}\right)\, ,
\end{multline}
where $K_n(x)$ is the modified Bessel function of the second kind. The other two  
integrals are
\begin{multline}\label{fl0(b,z)}
F'_0(z,\gamma',b)=\int^\infty_0 d\gamma~{J_0(b \sqrt{\gamma})
\over (\gamma+\gamma' +\kappa^2+z^2\frac{M^2}{4})^2} \\ 
=\frac{b\,K_1\left(b\,\sqrt{\gamma' +\kappa^2+z^2\tfrac{M^2}{4}}\right)}{\sqrt{\gamma' +\kappa^2+z^2\tfrac{M^2}{4}}}\, ,
\end{multline}
and
\begin{multline}\label{f1(b,z)}
F_1(z,\gamma',b)=\int^\infty_0 d\gamma~{\sqrt{\gamma}J_1(b \sqrt{\gamma})
\over [\gamma+\gamma' +\kappa^2+z^2\tfrac{M^2}{4}]^2} \\
= b K_0\left(b\,\sqrt{\gamma' +\kappa^2+z^2\tfrac{M^2}{4}}\right)\, .
\end{multline}
Notice that  $F_0$, $F'_0$ and $F_1$ depend upon $z^2$, and this allows one  to eliminate odd functions when integrating on
$z$ in Eq. \eqref{phibgm}.
The driving exponential fall-off of $F_0$, $F_0'$ and $F_1$  
in the asymptotic limit $b \rightarrow \infty$ comes from
 $K_{m}(x)$, which reads:
\begin{equation}\label{k1asymp}
K_{m}(x)|_{x\to\infty}\to\left( \frac{\pi}{2\,x} \right)^{\frac{1}{2}} {\rm e}^{-x} ~~.
\end{equation}
Hence,  the leading exponential behavior  in the  integrals \eqref{f0(b,z)}, 
\eqref{fl0(b,z)} and \eqref{f1(b,z)}
comes from values of ${\rm e}^{-b\,\sqrt{\gamma'+\kappa^2+z^2\tfrac{M^2}{4} }}$ 
(as seen from Eq.~\eqref{k1asymp}) with 
$\gamma'$ close to $0$ and $z\sim 0$, namely
\be\label{f2(b,z)}
F_0(z\sim 0,\gamma'\sim 0,b)|_{b\to\infty}=
\nonu
=b^{-1/2}F'_0((z\sim 0,\gamma'\sim 0,b)|_{b\to\infty}=\nonu =b^{-1}
F_1((z\sim 0,\gamma'\sim 0,b)|_{b\to\infty}\sim~{\rm e}^{-b\,\kappa}  ~~,
\ee
 with $\kappa$ given in Eq. \eqref{def_kappa}.
%
\end{document}